\documentclass[acmsmall,natbib,screen=true,anonymous=false,review=false]{acmart}

\PassOptionsToPackage{dvipsnames}{xcolor}

\usepackage{dsfont}
\usepackage{acronym}
\usepackage{amsmath}
\usepackage[linesnumbered,lined,ruled,commentsnumbered]{algorithm2e}
\usepackage{bbm}
\usepackage{bm}
\usepackage{booktabs}
\usepackage[skip=0pt]{caption}
\usepackage[T1]{fontenc}
\usepackage{graphicx}
\usepackage{hyperref}
\usepackage{listings}
\usepackage{multirow}
\usepackage{subcaption}
\usepackage{tabularx}
\usepackage[dvipsnames]{xcolor}
\usepackage{enumerate}
\usepackage[inline]{enumitem}
\usepackage{numprint}
\usepackage[normalem]{ulem}
\usepackage{mleftright}
\usepackage{soul}

\acrodef{GNN}{graph neural network}
\acrodef{IR}{information retrieval}
\acrodef{LTR}{learning to rank}
\acrodef{MC}{Markov chain}
\acrodef{NBR}{next basket recommendation}
\acrodef{PIF}{personal frequency information}
\acrodef{RNN}{recurrent neural network}
\acrodef{TREx}{two-step repetition-exploration}
\acrodef{TREx-Rep}{TREx with only repetition module}
\acrodef{SOTA}{state-of-the-art}

\newcommand{\RepR}{\mathit{RepR}}
\newcommand{\ExplR}{\mathit{ExplR}}

\newcommand{\tabincell}[2]{\begin{tabular}{@{}#1@{}}#2\end{tabular}}

\definecolor{paleyellow}{HTML}{FFEF77}
\definecolor{paleorange}{HTML}{FBB068}
\definecolor{paleblue}{HTML}{65B2FF}
\newcommand{\hlyellow}[1]{{\sethlcolor{paleyellow}\hl{#1}}}
\newcommand{\hlorange}[1]{{\sethlcolor{paleorange}\hl{#1}}}
\newcommand{\hlblue}[1]{{\sethlcolor{paleblue}\hl{#1}}}

\newcommand{\Better}[1]{\rlap{*}}

\author{Ming Li}
\orcid{0000-0001-7430-4961}
\affiliation{%
        \institution{AIRLab, University of Amsterdam}
        \city{Amsterdam}
        \country{The Netherlands}
}
\email{m.li@uva.nl}

\author{Sami Jullien}
\orcid{0000-0003-4507-6335}
\affiliation{%
        \institution{AIRLab, University of Amsterdam}
        \city{Amsterdam}
        \country{The Netherlands}
}
\email{s.jullien@uva.nl}

\author{Mozhdeh Ariannezhad}
\orcid{0000-0002-1113-8094}
\affiliation{%
        \institution{AIRLab, University of Amsterdam}
        \city{Amsterdam}
        \country{The Netherlands}
}
\email{m.ariannezhad@uva.nl}

\author{Maarten de Rijke}
\orcid{0000-0002-1086-0202} 
\affiliation{%
        \institution{University of Amsterdam}
        \city{Amsterdam}
        \country{The Netherlands}
}
\email{m.derijke@uva.nl}

\copyrightyear{2023}
\acmYear{2023}
\setcopyright{rightsretained}
\acmJournal{TOIS}
\acmPrice{}
\acmPrice{15.00}
\acmDOI{}
\acmISBN{}



\begin{document}

\title{A Next Basket Recommendation Reality Check}

\begin{abstract}
The goal of a \acl{NBR} system is to recommend items for the next basket for a user, based on the sequence of their prior baskets. 
We examine whether the performance gains of the \ac{NBR} methods reported in the literature hold up under a fair and comprehensive comparison.
To clarify the mixed picture that emerges from our comparison, we provide a novel angle on the evaluation of \ac{NBR} methods, centered on the distinction between repetition and exploration: the next basket is typically composed of previously consumed items (i.e., repeat items) and new items (i.e., explore items). 
We propose a set of metrics that measure the repetition/exploration ratio and performance of \ac{NBR} models.
Using these new metrics, we provide a second analysis of state-of-the-art \ac{NBR} models.
The results help to clarify the extent of the actual progress achieved by existing \ac{NBR} methods as well as the underlying reasons for any improvements that we observe. 
Overall, our work sheds light on the evaluation problem of \ac{NBR}, provides a new evaluation protocol, and yields useful insights for the design of models for this task.
\end{abstract}

\maketitle              

\acresetall


\section{Introduction}

Over the years, \ac{NBR} has received a considerable amount of interest from the research community \citep{fpmc,dream,bai2018attribute}. Baskets, or sets of items that are purchased or consumed together, are pervasive in many real-world services, with e-commerce and grocery shopping as prominent examples~\citep{hao, raeder}. 
Given a sequence of baskets that a user has purchased or consumed in the past, the goal of a \ac{NBR} system is to generate the basket of items that the user would like to purchase or consume next. 
Within a basket, items have no temporal order and are equally important. 
A key difference between \ac{NBR} and session-based or sequential item recommendations is that \ac{NBR} systems need to deal with multiple items in one set. 
Therefore, models designed for item-based recommendation are not fit for basket-based recommendation, and dedicated \ac{NBR} methods have been proposed~\citep{tifuknn, recency, dream, beacon, intnet, clea, sets2sets, sun2020timestamp2}. 

\subsection{Types of recommendation methods}
Over the years, we have seen the development of a wide range of recommendation methods.
\emph{Frequency-based} methods continue to play an important role as they are able to capture global statistics concerning popularity of items; this holds true for item-based recommendation scenarios as well as for \ac{NBR} scenarios.
Similarly, \emph{nearest neighbor-based} methods have long been used for both item-based and basket-based recommendation scenarios.
More recently, deep learning techniques have been developed to address sequential item recommendation problems, building on the capacity of deep learning-based methods to capture hidden relations and automatically learn representations~\citep{batmaz-2019-deep}. 
Recent years have also witnessed proposals to address different aspects of the \ac{NBR} task with deep learning-based methods, e.g., item-to-item relations~\citep{beacon}, cross-basket relations~\citep{dnntsp}, and noise within a basket~\citep{clea}. 

Recent analyses indicate that deep learning-based approaches may not be the best performing for all recommendation tasks and under all conditions~\citep{jannach-2020-why}.
For the task of generating a personalized ranked list of items, linear models and nearest neighbor-based approaches outperform deep learning-based methods~\citep{dacrema-2021-troubling}.
For sequential recommendation problems, deep learning-based methods may be outperformed by simple nearest neighbor or graph-based baselines~\citep{dacrema-2019-are}.
What about the task of \acl{NBR}? 
Here, the unit of retrieval --- a basket -- is more complex than in the recommendation scenarios considered in~\citep{dacrema-2019-are,dacrema-2021-troubling,jannach-2020-why}, with complex dependencies between items and baskets, across time, thus creating a potential for sophisticated representation learning-based approaches to \ac{NBR} to yield performance gains. 
In this paper, we take a closer look at the field to see if this is actually true.

\subsection{A new analysis perspective}
We find important gaps and flaws in the literature on \ac{NBR}.
These include weak or missing baselines, the use of different datasets in different papers, and of non-standard metrics.
We evaluate the performance of three families of state-of-the-art \ac{NBR} models (frequency-based, nearest neighbor-based, and deep learning-based), on three benchmark datasets, and find that no \ac{NBR} method consistently outperforms all other methods across all settings.

\begin{figure*}[t]
    \centering
    \includegraphics[width=0.9\linewidth]{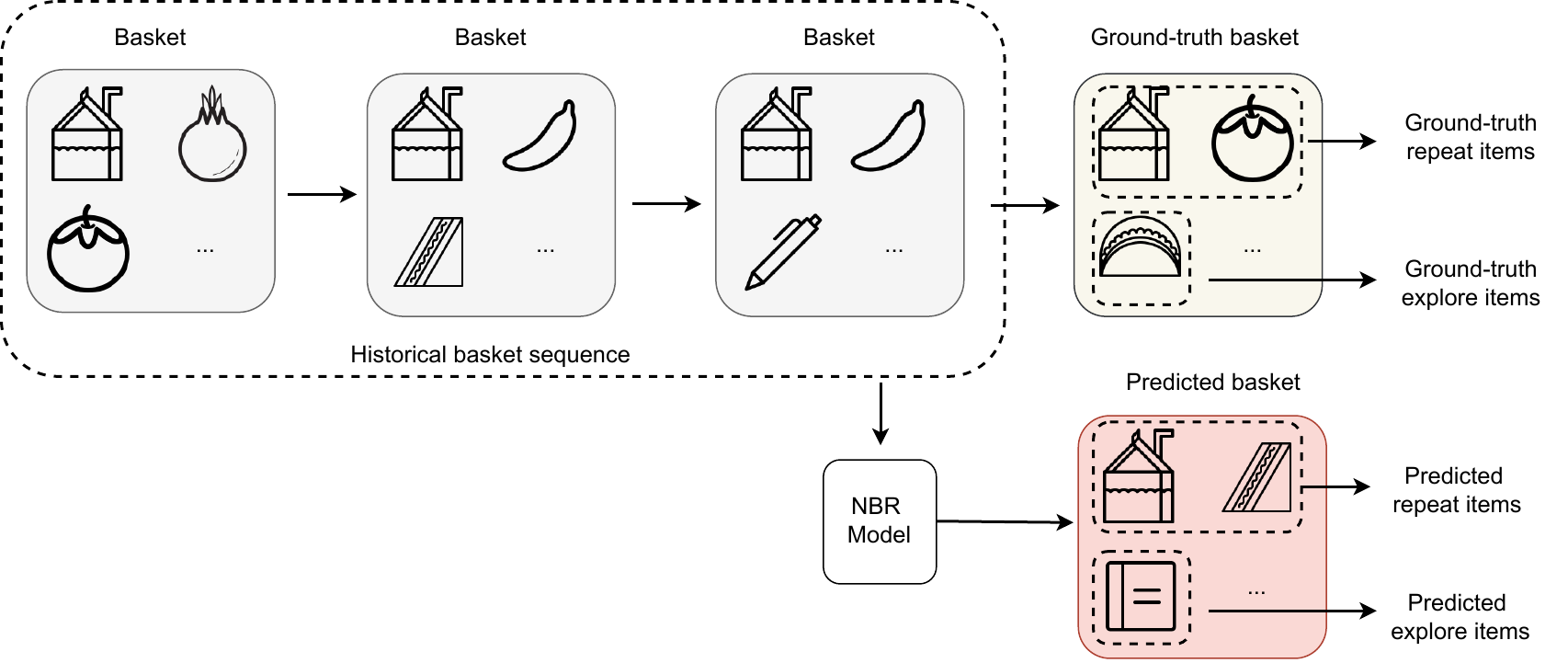}
    \caption{Four types of items in \acl{NBR}.}
    \label{fig:re-nbr}
\end{figure*}

Given these outcomes, we propose a more thorough analysis of the successes and failures of \ac{NBR} methods. 
As we show in Figure~\ref{fig:re-nbr}, baskets recommended in a \ac{NBR} scenario consist of \emph{repeat items} (items that the user has consumed before, in previous baskets) and \emph{explore items} (items that are new to the user).
The novelty of recommended items has been studied before, and related metrics have also been proposed~\citep{epc-epd}, but novelty-oriented metrics are not \ac{NBR} specific and only focus on one aspect, i.e., evaluating the novelty of the list of recommendations.
In order to improve our understanding of the relative performance of \ac{NBR} models, especially regarding repeat items and explore items, we introduce a set of task-specific metrics for \ac{NBR}.
Our newly proposed metrics help us understand which types of items are present in the recommended basket and assess the performance of \ac{NBR} models when proposing new items vs.\  already-purchased items. 

\subsection{Main findings}
Equipped with our newly proposed metrics for \ac{NBR}, we repeat our large-scale comparison of \ac{NBR} models and arrive at the following important findings:
\begin{itemize}
\item No \ac{NBR} method consistently outperforms all other methods across different datasets. 
\item All published methods are heavily skewed towards either repetition or exploration compared to the ground-truth, which might harm long-term engagement. 
\item There is a large \ac{NBR} performance gap between repetition and exploration; repeat item recommendation is much easier. 
\item In many settings, deep learning-based \ac{NBR} methods are outperformed by frequency-based baselines that fill a basket with the most frequent items in a user's history, possibly complemented with items that are most frequent across all users.
\item A bias towards repeat items accounts for most of the performance gains of recently published methods, even though many complex modules or strategies specifically target explore items. 
\item We propose a new protocol for evaluating \ac{NBR} methods, with a new frequency-based \ac{NBR} baseline as well as new metrics to assess the potential performance gains of \ac{NBR} methods.
\item Existing \acs{NBR} methods have different treatment effects on user performance and item exposure for users with different repetition ratios and items with different frequencies, respectively.
\end{itemize}

\noindent%
Overall, our work sheds light on the state-of-the-art of \ac{NBR}, provides suggestions to improve our evaluation methodology for \ac{NBR}, helps us understand the reasons underlying performance differences, and provides insights to inform the design of future \ac{NBR} models. 


\section{Related Work}
\if1
\subsection{Sequential and session-based recommendation}
Recurrent neural networks have shown strong performance in modeling sequential information, thus they have been widely used to learn the representation of the historical sequence behavior in session-based recommendation. GRU4Rec~\citep{gru4rec} first leverages GRUs to model user sequences and optimize a ranking-based loss for session-based recommendation. Later, an updated version, GRURec{+}~\citep{gru4recupdate} with a new ranking loss and sampling strategy has been proposed. The attention-based GRU(NARM)~\citep{narm} introduces an attention mechanism to make the recommendation model focus more on recent baskets. SASRec~\citep{sasrec} employs a self-attention-based method to capture the temporal dynamics of sequential recommendations in an efficient manner. 

In addition to RNN- and transformer-based models, several deep learning techniques are also applied to this area. Memory networks are applied by STAMP~\citep{stamp} to capture user's general and current interests. SR-GNN~\citep{srgnn} models session sequence as a graph and then uses a Graph Neural Network~\citep{gnn} to capture item transactions and learn an accurate item embedding. \citet{caser} propose a CNN based method to capture general interests and sequential patterns via vertical and horizontal filtering, whereafter NextItNet~\citep{nextitnet} introduces a generative model to improve the performance. Pre-trained model(BERT)~\citep{bert4rec} and Knowledge Graph are also applied to this case. 
\fi

\subsection{Reproducibility in information retrieval}
Reproducibility is a topic that has been at the center of \ac{IR} research for many years. 
The \emph{mechanics} of reproducibility have been a constant factor since the early days of community benchmarking~\citep{voorhees-2005-trec}, resulting in a large number of datasets and metrics.
Artifact badging is a matter of ongoing and active interest~\citep{ferro-2018-sigir} as are ways to objectively quantify to what extent a system-oriented \ac{IR} experiment has been replicated or reproduced~\citep{breuer-2020-how}. 

Asking \emph{which lessons hold up} under closer scrutiny is not new either in \ac{IR}. 
Papers of this type have been written for query performance prediction~\cite{hauff-2008-survey}, ranking~\citep{armstrong-2009-improvements}, learning to rank~\citep{tax-2015-cross-benchmark}, search result diversification~\citep{akcay-2017-additivity}, online learning to rank~\citep{oosterhuis-2019-optimizing}, question answering~\citep{crane-2018-questionable}, and neural rankers~\citep{yang-2019-critically,lin-2019-neural}.
We are particularly interested in this ``which lessons hold up'' aspect of reproducibility in the context of recommender systems.
\citet{dacrema-2019-are,dacrema-2021-troubling} and \citet{jannach-2020-why} have recently examined the relative strength of deep learning-based methods for item recommendation, both in a traditional static setting and in a sequential setting.
In contrast, we consider the task of \acfi{NBR} and assess the relative merits of deep learning-based methods for this task.

Some \ac{NBR} methods \cite{dream, bai2018attribute, beacon, clea, intnet} only compare with previous (deep) learning-based methods and avoid comparing with frequency-based baselines that recommend the $k$ most frequent items in a users’ historical records as the next basket. 
In several cases, recent publications on \ac{NBR} omit comparisons to other recent methods~\citep{recency, tifuknn, dnntsp, clea}. 
In~\cite{clea}, sampled metrics \cite{sasrec} are used to evaluate the performance even though this is not encouraged~\citep{sampledmetrics}.
As our systematic comparisons below show, not all previously published lessons on deep learning-based \ac{NBR} methods hold up.

\subsection{Next basket recommendation}
The \ac{NBR} problem has been studied for many years. 
As we explain in Section~\ref{section:dataset-and-methods} below, we analyze the performance of three families of \ac{NBR} methods.
First, we consider frequency-based methods; in different configurations they have been considered as baselines in most prior work on \ac{NBR} that we are aware of.
Second are nearest neighbor-based methods. TIFUKNN~\citep{tifuknn} and UP-CF@r~\citep{recency} model temporal patterns over frequency information and then combine with neighbor information or user-wise collaborative filtering.
Third are deep learning-based methods.
Such methods often have a strong focus on learning representations of baskets. 
Early precursors include the factorizing personalized Markov chains (FPMC)~\citep{fpmc}, which leverage matrix factorization (MF) and Markov chains to model users' general interest and basket transition relations, and hierarchical representation models (HRMs)~\citep{hrm}, which apply aggregation operations to learn a hierarchical representation of baskets; these two MC-based methods only capture local short-term relations between adjacent baskets. 
In contrast, RNNs have been used for the \ac{NBR} task to learn long-term trends by modeling the whole basket sequence. 
For instance, DREAM~\citep{dream} uses max/avg pooling to encode baskets. 
Sets2Sets~\citep{sets2sets} adapts an attention mechanism and adds frequency information to improve performance. 
Some methods~\citep{beacon, intnet} consider item relations to obtain a better representation. 
\citet{dnntsp} argue that item-item relations between baskets are important and leverage GNNs to capture these relations; the authors also use a self-attention mechanism to learn temporal dependencies between baskets. 
Some methods~\citep{bai2018attribute, wang2019timestamp1, sun2020timestamp2, nbrdiversity, anda} exploit auxiliary information, including product categories, amounts, prices, and explicit time stamps; for the sake of a fair comparison, we omit these from our reproducibility study.

What we add on top of prior work is not yet another \ac{NBR} method but a systematic comparison under the same experimental conditions across multiple datasets as well as an analysis of the relative performance of state-of-the-art \ac{NBR} methods in terms of repetition and exploration, which helps to explain the observed performance differences.


\section{Datasets and Methods}
\label{section:dataset-and-methods}

\subsection{Datasets}
In order to ensure the reproducibility of our study, we conduct our experiments on three publicly available real-world datasets:
\begin{itemize}
    \item \textbf{TaFeng} -- contains four months of shopping transactions collected from a Chinese grocery store. All products purchased on the same day by the same user are treated as a basket.\footnote{\url{https://www.kaggle.com/chiranjivdas09/ta-feng-grocery-dataset}} 
    \item \textbf{Dunnhumby} -- covers two years of household-level transactions at a retailer. All products bought by the same user in the same transaction are treated as a basket. We use the first two months of the data in our experiments.\footnote{\url{https://www.dunnhumby.com/source-files/}} 
    \item \textbf{Instacart} -- contains over three million grocery orders of Instacart users. We treat all items purchased by the same user in the same order as a basket.\footnote{\url{https://www.kaggle.com/c/instacart-market-basket-analysis/data}}
\end{itemize}

\noindent%
Following previous work~\citep{dream, beacon, tifuknn, sets2sets, dnntsp}, we also employ a sampling strategy instead of using the whole dataset. In each dataset, users with a basket size between 3 and 50 are sampled to conduct experiments. We also remove rare and unpopular items and the remainder covers more than 95\% of the interactions. 
A \emph{ground truth basket} is a basket that we aim to predict or recommend, and the last basket of a sequence or purchased baskets is regarded as the ground truth basket.
The \emph{repetition ratio} and \emph{exploration ratio} are calculated based on the ground truth basket as the proportion of \emph{repeat items} and \emph{explore items} in the ground truth baskets, respectively.
The statistics of the processed datasets are summarized in Table~\ref{dataset}.

For our experiments, we split every dataset across users like previous works~\citep{dnntsp,sets2sets,tifuknn}, i.e., 72\% users for training, 8\% users for validation, and 20\% users for the test set. The training users, validation users, and test users are totally independent from each other.
We repeat the dataset split five times for independent experiments.
Note that we did not use an absolute timestamp splitting strategy (i.e., splitting the dataset into several sub-datasets according to real time ranges) for the following reasons:
\begin{enumerate}
  \item As a reality check paper, we decided to follow the widely used data splitting strategy in the existing \acs{NBR} methods.
  \item We checked three datasets characteristics and found that users' baskets are spread out across diverse time ranges and the average basket lengths are limited, so splitting using an absolute timestamp would be likely to break a large number of basket sequences into very short sequences, which cannot be used for effectively training or evaluating NBR methods.
  \item As the paper focuses on the analysis from a repetition and exploration perspective, we acknowledge that global trends are likely to influence the repetition ratio of the users, therefore we use Section~\ref{sec: treatment-users} to analyze at a fine-grained level, i.e., the treatment on users with different repetition ratios.
\end{enumerate}

\begin{table}
\centering
  \caption{Statistics of the processed datasets.}
  \label{dataset}
  \setlength{\tabcolsep}{2.5pt}
  \begin{tabular}{lcccccc}
    \toprule
    Dataset & \#Items & \#Users & \tabincell{c}{Avg. \\basket\\ size} & \tabincell{c}{Avg. \\\#baskets\\ per user}  & \tabincell{c}{Repeat \\ ratio} & \tabincell{c}{Explore \\ ratio}\\
    \midrule
    TaFeng & 11,997 & 13,858 & 6.27 & \phantom{1}6.58 & 0.188 & 0.812 \\
    Dunnhumby & \phantom{0}3,920 & 22,530 & 7.45 & \phantom{1}9.53 & 0.409 & 0.591\\
    Instacart & 13,897 & 19,435 & 9.61 & 13.21 & 0.597 & 0.403\\
    \bottomrule
  \end{tabular}
\end{table}

\subsection{Baseline methods and reproducible methods}
\label{subsection:baselines}
We follow the same strategy as \cite{dacrema-2019-are} to collect relevant and reproducible \ac{NBR} papers. 
Specifically, we include papers in our analysis that have been published during the last five years (i.e., from 2016 to 2021) in one of the following conferences: KDD~\citep{dnntsp, sets2sets, sun2020timestamp2}, SIGIR~\citep{dream,bai2018attribute, tifuknn}, IJCAI~\citep{beacon}, AAAI~\citep{intnet}, RecSys~\citep{anda} and UMAP~\citep{recency}.
Papers targeting \ac{NBR}~\citep{dream, bai2018attribute, beacon, tifuknn, clea, recency, intnet} and sequential set prediction~\citep{sets2sets, dnntsp, sun2020timestamp2} are considered to be relevant papers. 
For a fair comparison, methods~\citep{bai2018attribute, anda, sun2020timestamp2} using auxiliary information other than item-basket sequences are not included in this paper. 
Like~\citep{dacrema-2019-are}, we consider a paper to be reproducible if they meet the following criteria: 
\begin{enumerate}
  \item A working version of the \emph{source code} is publicly available\footnote{We first check whether the paper provides a link to their code; if not, we search GitHub using the title of the paper.} or the code has to be modified in minimal ways to work correctly.\footnote{We re-implement the Dream algorithm~\citep{dream}.}
  \item At least one \emph{dataset} used in the original paper is available.
\end{enumerate}
Through this selection process, we end up with eight relevant representative papers~\citep{dream, beacon, intnet, sets2sets, dnntsp, tifuknn, recency, clea}, and seven of them are considered to be reproducible methods in our analysis~\citep{dream, beacon, sets2sets, dnntsp, tifuknn,recency, clea}.\footnote{We do not include FPMC~\citep{fpmc} in our paper for the following two reasons: first, it will break the selection criteria~\citep{dacrema-2019-are} we employ in this paper; second, FPMC is among the worst performing method in all \acs{NBR} related papers~\citep{dream, beacon, tifuknn, clea} recently.}
Of the seven reproducible methods, four methods~\citep{dnntsp, tifuknn, recency, clea} have been published during the last two years, and have not been compared with each other.

As simple, yet effective baselines we include two widely known frequency-based methods, i.e., \emph{global top-frequency} (G-TopFreq) and \emph{personal top-frequency} (P-TopFreq), which are often shown as simple baselines in recommendation tasks. Surprisingly, we find that 3 of the 5 deep-learning based methods that we consider only compare with the global top-frequency baseline G-TopFreq~\citep{dream,beacon, clea}, but do not compare to the personal top-frequency baseline P-TopFreq, which is known to have higher performance in general~\cite{sets2sets, dnntsp, tifuknn,recency}.

There is an important limitation of the \emph{personal top-frequency} method (P-TopFreq) w.r.t.\ the basket size in a \acs{NBR} setting that is ignored in previous work.
P-TopFreq can only recommend items from the past transactions of a user, which means that it might not be able to fully make use of the available basket slots like other methods, and this may lead to an unfair comparison.
We analyze the percentage of basket slots used for P-TopFreq, and Table~\ref{dataset: rep_slots} shows the results. 
To address this limitation, we propose a simple combination of G-TopFreq and P-TopFreq as an additional baseline, called \emph{GP-TopFreq}: GP-TopFeq first uses P-TopFreq to fill a basket, and then uses G-TopFeq to fill any remaining slots.

\begin{table}
  \centering
    \caption{Percentage of the basket slots used for P-TopFreq.}
    \label{dataset: rep_slots}
    \setlength{\tabcolsep}{2.5pt}
    \begin{tabular}{cccc}
      \toprule
      & \multicolumn{3}{c}{Dataset} \\
      \cmidrule{2-4}
      Basket size& TaFeng & Dunnhumby &  Instacart\\
      \midrule
      10 &92.39\% & 95.66\% & 96.70\%  \\
      20 & 79.71\% & 87.98\% & 90.48\% \\
      \bottomrule
    \end{tabular}
  \end{table}

\subsubsection{Frequency-based baselines}
\begin{itemize}
    \item \textbf{G-TopFreq} -- uses the $k$ most popular items in the dataset to form the recommended next basket. It is widely used in recommendation systems due to its effectiveness and simplicity.
    \item \textbf{P-TopFreq} -- a personalized top frequency method that treats the most frequent $k$ items in the users' historical records as the next basket. This method only has repeat items in the prediction. 
    \item \textbf{GP-TopFreq} -- a combination of P-TopFreq and G-TopFreq to make full use of the available basket slots.
\end{itemize}

\subsubsection{Nearest neighbor-based methods}
\begin{itemize}
    \item \textbf{TIFUKNN} -- models the temporal dynamics of frequency information of users' past baskets to introduce personalized frequency information~(PIF), then uses a nearest neighbor-based method on PIF~\citep{tifuknn}.
    \item \textbf{UP-CF@r} -- a combination of recency-aware, user-wise popularity and user-wise collaborative filtering. The recency of shopping behavior is considered in this method~\citep{recency}.
\end{itemize}

\subsubsection{Deep learning-based methods}
\begin{itemize}
    \item \textbf{Dream} -- the first deep learning-based method that models users' global sequential basket history for NBR. It uses a pooling strategy to generate basket representations, which are then fed into an RNN to learn user representations and predict the corresponding next set of items~\citep{dream}.
    \item \textbf{Sets2Sets} -- uses a pooling operation to get basket embeddings and an attention mechanism to learn a user's representation from their past interactions. Furthermore, item frequency information is adopted to improve performance~\citep{sets2sets}.
    \item \textbf{DNNTSP} -- leverages a GNN and self-attention techniques. It encodes item-item relations via a graph and employs a self-attention mechanism to capture temporal dependencies of users' basket sequence~\citep{dnntsp}.
    \item \textbf{Beacon} -- an RNN-based method that  encodes the basket considering the incorporating information on pairwise correlations among items~\citep{beacon}. 
    \item \textbf{CLEA} -- an RNN-based method that uses a contrastive learning model to automatically extract items relevant to the target items and generates the representation via a GRU-based encoder~\citep{clea}.
\end{itemize}

\subsection{Implementation details}
For deep learning-based methods~\cite{dream, sets2sets, dnntsp, beacon, clea}, we strictly follow the hyper-parameter setting and tuning strategy in their respective paper or related GitHub repository. 
Following the same strategy as \cite{dacrema-2019-are}, we use the suggested best parameters in TIFUKNN~\citep{tifuknn} to achieve its best performance.
For UP-CF@r, the recency window is tuned on \{1, 5, 10, 50\}, the locality is tuned on \{1, 20, 50, 100\}, the asymmetry is tuned on \{0, 0.25, 0.5, 0.75, 1.0\}.
We perform a grid search on the validation dataset to tune hyper-parameters and select the best model for testing.
For all methods, we rely as much as possible on the original source code and construct a pipeline to perform experiments. We share the code and data used in our experiments online.\footnote{Code available at \url{https://github.com/liming-7/A-Next-Basket-Recommendation-Reality-Check}.}


\section{Performance Comparison Using Conventional Metrics}
\label{section:overall}

\begin{table}[t]
	\centering
	\caption{Notation used in the paper.}
	\setlength{\tabcolsep}{2pt}
	\label{tab:notations}
	\begin{tabular}{lp{12cm}}
		\toprule
		Symbol & Description \\
		\midrule
		$U$ & Set of all users \\
    $U_g$ & Set of users in group $g$\\
		$I$ & Set of all items \\
		$u$ & A single user in $U$\\
		$i$ & A single item in $I$ \\
		$S_{j}$ & Sequence of historical baskets for $u_{j}$ \\
		$B_{j}^{t}$ & $t$-th basket in  $S_{j}$, a set of items $i \in I$ \\
		$T_{u_{j}}$ & Target/ground truth basket for $u_{j}$ that we aim to predict\\
		$T_{u_{j}}^\mathit{rep}$ & Set of \emph{repeat items} in the ground truth basket $T_{u_{j}}$ for $u_{j}$\\
		$T_{u_{j}}^\mathit{expl}$ & Set of \emph{explore items} in the ground truth basket $T_{u_{j}}$ for $u_{j}$\\
		$P_{u_{j}}$ & Predicted basket for $u_{j}$  \\
		$P_{u_{j}}^\mathit{rep}$ & Set of \emph{repeat items} in the predicted basket $P_{u_{j}}$ for $u_{j}$\\
		$P_{u_{j}}^\mathit{expl}$ & Set of \emph{explore items} in the predicted basket $P_{u_{j}}$ for $u_{j}$\\
		$I_{j,t}^\mathit{rep}$ &Repeat items for $u_j$ at timestamp $t$; set of items that have appeared in $u_j$'s baskets up to  timestamp $t$\\
		$I_{j,t}^\mathit{expl}$ &Explore items for $u_j$ at timestamp $t$; set of items that 
		have not appeared in $u_j$'s baskets up to  timestamp $t$\\ 
		\bottomrule
	\end{tabular}
\end{table}

\subsection{Conventional \ac{NBR} metrics}
To analyze the performance of \ac{NBR} methods, we first consider three conventional metrics: recall, normalized discounted cumulative gain (NDCG), and personalized hit ratio (PHR), all of which are commonly used in previous \ac{NBR} studies~\citep{tifuknn,sets2sets,dnntsp}. 
We do not consider the F1 and Precision metrics in this paper, since we focus on the basket recommendation with a fixed basket size $K$, which means the Precision@$K$ and F1@$K$ are proportional to Recall@$K$ for each user. F1@$K$ and Precision@$K$ are more suitable for \acs{NBR} with a dynamic basket size for each user.

Recall measures the ability to find all relevant items and is calculated as follows:
\begin{equation}
Recall@K(u_j) = \frac{\left | P_{u_j}\cap T_{u_j} \right |}{\left |T_{u_j}\right|},
\end{equation}
where $P_{u_j}$ is the predicted basket with $K$ recommended items and $T_{u_j}$ is the ground truth basket for user $u_j$. 
The average recall score of all users is adopted as the recall performance.

NDCG is a ranking quality measurement metric, which takes item order into consideration and it is calculated as follows, for a user $u \in U$ and its ground truth basket $T_u$:
\begin{equation}
\mathrm{NDCG}@K(u_j) = \frac{\sum_{k=1}^Kp_k/\log_2(k+1)}{\sum_{k=1}^{\min(K, |T_{u_j}|)} 1/\log_2(k+1)},
\end{equation}
where $p_k$ equals 1 if $P_{u_j}^k\in T_u$, otherwise $p_k=0$.  $P_u^k$ denotes the $k$-th item in the predicted basket $P_u$. The average score across all users is the NDCG performance of the algorithm.

PHR focuses on user level performance and calculates the ratio of predictions that capture at least one item in the ground truth basket as follows:
\begin{equation}
\mathrm{PHR}@K = \frac{\sum_{j=1}^N\varphi(P_{u_j}, T_{u_j})}{N},
\end{equation}

where $N$ is the number of test users, and $\varphi(P_{u_j}, T_{u_j})$ returns 1 when $P_{u_j} \cap T_{u_j} \neq \emptyset$, otherwise returns 0.

\subsection{Results with conventional \ac{NBR} metrics}

\begin{table*}[!t]
  \caption{Overall performance comparison of frequency-based, nearest neighbor-based, and deep learning-based \ac{NBR} methods. \protect\hlyellow{Highlights} indicate the highest score per basket size and metric. We write * to indicate that the highest score for a given basket size and metric is significantly better than the second highest score (paired t-test, p-value $<0.05$). }
  \label{tab:conventional}
  \scalebox{0.9}{%
  \begin{tabular}{ll ccc ccc ccc}
    \toprule
    \multicolumn{2}{l}{Size} & \multicolumn{3}{c}{10} & \multicolumn{3}{c}{20}\\
    \cmidrule(r){3-5}
    \cmidrule(r){6-8}
    Dataset & Methods & Recall & NDCG &  PHR & Recall & NDCG & PHR \\
    \midrule
    \multirow{10}{*}{\rotatebox[origin=c]{90}{TaFeng}}& G-TopFreq & 0.0831 \tiny{(0.0018)} & 0.0864 \tiny{(0.0017)}& 0.2498 \tiny{(0.0024)} & 0.1114 \tiny{(0.0029)} & 0.0961 \tiny{(0.0018)} & 0.3311 \tiny{(0.0006)} \\
    &P-TopFreq & 0.1069 \tiny{(0.0023)} & 0.0955 \tiny{(0.0019)} & 0.3473 \tiny{(0.0033)} & 0.1395 \tiny{(0.0026)} & 0.1096 \tiny{(0.0019)} & 0.4329 \tiny{(0.0038)}\\
    &GP-TopFreq & 0.1211 \tiny{(0.0031)} & 0.1015 \tiny{(0.0023)} & 0.3691 \tiny{(0.0043)} & 0.1693 \tiny{(0.0031)} & 0.1208 \tiny{(0.0022)} & 0.4834 \tiny{(0.0040)}\\
    \cmidrule{2-8}
    &UP-CF@r & 0.1249 \tiny{(0.0027)} & 0.1104 \tiny{(0.0019)} & 0.3983 \tiny{(0.0035)} & 0.1694 \tiny{(0.0034)} & 0.1280 \tiny{(0.0021)} & 0.4877 \tiny{(0.0048)}\\
    &TIFUKNN & 0.1251 \tiny{(0.0033)} & 0.1016 \tiny{(0.0014)} & 0.3852 \tiny{(0.0029)} & 0.1817 \tiny{(0.0037)} & 0.1232 \tiny{(0.0016)} & 0.5043 \tiny{(0.0035)}\\    
    \cmidrule{2-8}
    &Dream& 0.1134 \tiny{(0.0023)} & 0.1022 \tiny{(0.0018)} & 0.3035 \tiny{(0.0024)} & 0.1463 \tiny{(0.0034)} & 0.1149 \tiny{(0.0021)} & 0.3905 \tiny{(0.0039)} \\
    &Beacon & 0.1139 \tiny{(0.0032)} & 0.1033 \tiny{(0.0023)} & 0.3055 \tiny{(0.0044)} & 0.1475 \tiny{(0.0026)} & 0.1154 \tiny{(0.0020)} & 0.4002 \tiny{(0.0024)} \\
    &CLEA & 0.1184 \tiny{(0.0038)} & 0.1046 \tiny{(0.0030)} & 0.3076 \tiny{(0.0044)} & 0.1477 \tiny{(0.0035)} & 0.1165 \tiny{(0.0028)} & 0.3957 \tiny{(0.0037)}\\
    &Sets2Sets & 0.1349 \tiny{(0.0034)} & 0.1124 \tiny{(0.0025)} & 0.4080 \tiny{(0.0058)} & 0.1904 \tiny{(0.0021)} & 0.1342 \tiny{(0.0020)} & 0.5261 \tiny{(0.0041)}\\
    &DNNTSP & \hlyellow{0.1526 \tiny{(0.0034)}}\Better & \hlyellow{0.1314 \tiny{(0.0022)}}\Better & \hlyellow{0.4460 \tiny{(0.0044)}}\Better & \hlyellow{0.2077 \tiny{(0.0044)}}\Better & \hlyellow{0.1532 \tiny{(0.0025)}}\Better & \hlyellow{0.5496 \tiny{(0.0038)}}\Better{} \\
    \midrule
    \multirow{10}{*}{\rotatebox[origin=c]{90}{Dunnhumby}}& G-TopFreq & 0.0982 \tiny{(0.0009)} & 0.1050 \tiny{(0.0010)} & 0.4621 \tiny{(0.0033)} & 0.1264 \tiny{(0.0006)} & 0.1192 \tiny{(0.0008)} & 0.5314 \tiny{(0.0030)} \\
    &P-TopFreq & 0.2319 \tiny{(0.0013)} & 0.2340 \tiny{(0.0012)} & 0.6559 \tiny{(0.0017)} & 0.3030 \tiny{(0.0012)} & 0.2695 \tiny{(0.0011)} & 0.7173 \tiny{(0.0017)}\\
    &GP-TopFreq & 0.2356 \tiny{(0.0013)} & 0.2358 \tiny{(0.0013)} & 0.6649 \tiny{(0.0019)} & 0.3141 \tiny{(0.0012)} & 0.2743 \tiny{(0.0011)} & 0.7372 \tiny{(0.0026)}\\
    \cmidrule{2-8}
    &UP-CF@r & \hlyellow{0.2428 \tiny{(0.0006)}} & \hlyellow{0.2468 \tiny{(0.0012)}} & 0.6747 \tiny{(0.0024)} & 0.3185 \tiny{(0.0009)} & \hlyellow{0.2848 \tiny{(0.0013)}} & 0.7339 \tiny{(0.0014)}\\
    &TIFUKNN & 0.2396 \tiny{(0.0008)} & 0.2408 \tiny{(0.0011)} & \hlyellow{0.6761 \tiny{(0.0022)}} & 0.3191 \tiny{(0.0015)} & 0.2799 \tiny{(0.0012)} & 0.7407 \tiny{(0.0030)}\\    
    \cmidrule{2-8}
    &Dream& 0.0950 \tiny{(0.0008)} & 0.1036 \tiny{(0.0008)} & 0.4586 \tiny{(0.0040)} & 0.1300 \tiny{(0.0013)} & 0.1205 \tiny{(0.0011)} & 0.5395 \tiny{(0.0033)}\\
    &Beacon & 0.0995 \tiny{(0.0009)} & 0.1066 \tiny{(0.0011)} & 0.4700 \tiny{(0.0037)} & 0.1354 \tiny{(0.0009)} & 0.1241 \tiny{(0.0010)} & 0.5506 \tiny{(0.0025)}\\
    &CLEA & 0.1552 \tiny{(0.0010)} & 0.1732 \tiny{(0.0010)} & 0.5541 \tiny{(0.0038)} & 0.1866 \tiny{(0.0012)} & 0.1864 \tiny{(0.0007)} & 0.6273 \tiny{(0.0029)}\\
    &Sets2Sets & 0.1691 \tiny{(0.0023)} & 0.1473 \tiny{(0.0015)} & 0.5802 \tiny{(0.0053)} & 0.2552 \tiny{(0.0018)} & 0.1880 \tiny{(0.0015)} & 0.6893 \tiny{(0.0046)}\\
    &DNNTSP & 0.2404 \tiny{(0.0007)} & 0.2430 \tiny{(0.0010)} & 0.6767 \tiny{(0.0022)} & \hlyellow{0.3242 \tiny{(0.0004)}}\Better & 0.2839 \tiny{(0.0007)} & \hlyellow{0.7427 \tiny{(0.0016)}}\\
    \midrule
    \multirow{10}{*}{\rotatebox[origin=c]{90}{Instacart}}& G-TopFreq & 0.0710 \tiny{(0.0003)} & 0.0811 \tiny{(0.0001)}& 0.4542 \tiny{(0.0010)}& 0.0990 \tiny{(0.0001)} & 0.0962 \tiny{(0.0002)} & 0.5248 \tiny{(0.0017)} \\
    &P-TopFreq & 0.3260 \tiny{(0.0008)} & 0.3378 \tiny{(0.0007)} & 0.8449 \tiny{(0.0018)} & 0.4307 \tiny{(0.0008)} & 0.3939 \tiny{(0.0002)} & 0.8957 \tiny{(0.0019)}\\
    &GP-TopFreq & 0.3269 \tiny{(0.0008)} & 0.3383 \tiny{(0.0007)} & 0.8463 \tiny{(0.0018)} & 0.4354 \tiny{(0.0007)} & 0.3961 \tiny{(0.0003)} & 0.9011 \tiny{(0.0017)}\\
    \cmidrule{2-8}
    &UP-CF@r & 0.3506 \tiny{(0.0007)} & 0.3631 \tiny{(0.0007)} & \hlyellow{0.8652 \tiny{(0.0020)}} & 0.4591 \tiny{(0.0008)} & 0.4222 \tiny{(0.0005)} & 0.9079 \tiny{(0.0012)}\\
    &TIFUKNN & \hlyellow{0.3601 \tiny{(0.0015)}}\Better & \hlyellow{0.3721 \tiny{(0.0008)}}\Better & 0.8642 \tiny{(0.0005)} & \hlyellow{0.4709 \tiny{(0.0015)}}\Better & \hlyellow{0.4323 \tiny{(0.0003)}}\Better & \hlyellow{0.9097 \tiny{(0.0011)}}\\    
    \cmidrule{2-8}
    &Dream& 0.0712 \tiny{(0.0004)} & 0.0805 \tiny{(0.0005)} & 0.4551 \tiny{(0.0008)} & 0.0997 \tiny{(0.0006)} & 0.0957 \tiny{(0.0007)} & 0.5304 \tiny{(0.0027)} \\
    &Beacon & 0.0734 \tiny{(0.0003)} & 0.0838 \tiny{(0.0003)} & 0.4628 \tiny{(0.0006)} & 0.1050 \tiny{(0.0006)} & 0.1009 \tiny{(0.0004)} & 0.5462 \tiny{(0.0022)}\\
    &CLEA & 0.1221 \tiny{(0.0014)} & 0.1449 \tiny{(0.0016)} & 0.5603 \tiny{(0.0053)} & 0.1514 \tiny{(0.0045)} & 0.1592 \tiny{(0.0019)} & 0.6347 \tiny{(0.0094)}\\
    &Sets2Sets & 0.2125 \tiny{(0.0013)} & 0.1923 \tiny{(0.0019)} & 0.7185 \tiny{(0.0040)} & 0.3077 \tiny{(0.0040)} & 0.2402 \tiny{(0.0020)} & 0.8284 \tiny{(0.0040)}\\
    &DNNTSP & 0.3330 \tiny{(0.0003)} & 0.3412 \tiny{(0.0004)} & 0.8525 \tiny{(0.0011)} & 0.4423 \tiny{(0.0005)} & 0.4000 \tiny{(0.0005)} & 0.9042 \tiny{(0.0003)}\\
    \bottomrule
  \end{tabular}}
\end{table*}
Performance results for the conventional \ac{NBR} metrics are shown in Table~\ref{tab:conventional}.
The performance of different methods varies across datasets; there is no method that consistently outperforms all other methods, independent of dataset and basket size. 
This calls for a further analysis of the factors impacting performance, which we conduct in the next section.

Among the frequency-based baselines, P-TopFreq outperforms G-TopFreq in all scenarios, which indicates that personalization improves the \ac{NBR} performance. P-TopFreq can only recommend items that have appeared in a user's previous baskets. 
As pointed out in Section~\ref{subsection:baselines}, the number of repeat items of a user may be smaller than the basket size, which means there might be empty slots in a basket recommended by P-TopFreq. 
Despite this limitation, P-TopFreq is a competitive \ac{NBR} baseline. 
GP-TopFreq makes full use of the available basket slots by filling any slots with top ranked items suggested by G-TopFreq. 
GP-TopFreq outperforms P-TopFreq with no surprise, and, as expected, the difference shrinks as the repetition ratio of the dataset increases. 
For future fair comparisons, we believe that GP-TopFreq should be the baseline for every \ac{NBR} method to compare with, especially in high exploration scenarios, to be able to determine what value is added beyond simple frequency-based recommendations.

As to the nearest neighbor-based methods, we see that TIFU\-KNN and UP-CF@r have a similar performance across different scenarios and outperform all frequency-based baselines. 
The two methods are similar in the sense that both model temporal information, combined with a user-based nearest neighbor method. 
Their performance in a high exploration scenario is lower than several deep learning-based methods (i.e., the TaFeng dataset), but on the Dunnhumby and Instacart datasets, which have a relatively low exploration ratio, they are among the best performing methods.

Most of the deep learning-based methods outperform G-TopFreq, which is the only frequency-based baseline considered in many papers. 
Surprisingly, P-TopFreq and GP-TopFreq achieve a highly competitive performance and outperform four deep learning-based methods (i.e., Dream, Beacon, CLEA and Sets2sets), by a large margin in the Dunnhumby and Instacart datasets, where the improvements in terms of $Recall@10$ range from 35.8\% to 141.9\% and from 53.6\% to 353.3\%, respectively.
Moreover, the proposed GP-Topfreq baseline outperforms the deep learning-based Beacon, Dream and CLEA algorithm on the TaFeng dataset, the scenario with a high exploration ratio. 
Of the deep learning-based methods, DNNTSP is the only one to have a consistently high performance in all scenarios. 

\subsection{Upshot}
\label{sec: upshot-overall}
Based on the above experiments and analysis, we conclude that the choice of dataset plays an important role in evaluating the performance of \ac{NBR} methods, and no state-of-the-art \ac{NBR} method is able to consistently achieve the best performance across datasets.

Several deep learning-based \ac{NBR} methods~\citep{dream, beacon, clea} aim to learn better performing representations by capturing long-term temporal dependencies, denoising, etc. They do indeed outperform the G-TopFreq baseline, but many are inferior to the P-TopFreq baseline, especially in datasets with a relatively high repetition ratio.
The proposed GP-TopFreq baseline in some sense ``re-calibrates'' the improvements reported for recently introduced complex, deep learning-based \acs{NBR} methods; compared to the simple GP-TopFreq baseline, their improvements are modest or even absent.\footnote{According to our analyses, performance differences that were reported in previous work still stand. However, the set of baselines used for comparison in previous work is too limited.}

So far we have used conventional metrics to examine the performance of \ac{NBR} methods. 
To account for the findings reported in this section and provide insights for future \acs{NBR} method development, we will now consider additional metrics.


\section{Performance w.r.t. Repetition and Exploration}
In order to understand which factors contribute to the overall performance of a \ac{NBR} method, we dive deeper into the basket components from a repetition and exploration perspective.

\subsection{Metrics for repetition vs.\ exploration}
We propose several metrics that are meant to capture repetition and exploration aspects of a basket.
First, we adopt widely used definitions of repetition and exploration in the recommender systems literature~\citep{repeatagain,repeatconsumtion1,repeatconsumtion2,epc-epd} to define what constitutes a repeat item and an explore item in the context of \acs{NBR}.
Specifically, an item $i^r$ is considered to be a \emph{repeat item} for a user $u_j$ if it appears in the sequence of the user's historical baskets $S_j$, that is, if \smash{$i^r \in I_{j, t}^\mathit{rep} = {B_j^1 \cup B_j^2 \cup \cdots \cup B_j^t} $}. 
Otherwise, the item is an \emph{explore item}, denoted as \smash{$i^e \in I_{j, t}^\mathit{expl} = I - I_{j, t}^\mathit{rep}$}. 
We write \smash{$P_{u_j}=P_{u_j}^\mathit{rep} \cup P_{u_j}^\mathit{expl}$} for the predicted next basket $B_j^{t+1}$, which is the union of \emph{repeat items} \smash{$P_{u_j}^\mathit{rep}$} and \emph{explore items} \smash{$P_{u_j}^\mathit{expl}$}.
As an edge case, a basket may consist of repeat or explore items only. 

\emph{Novelty of recommendation} is a concept that is similar to, but different from, the notion of exploration that we use in this paper. Several novelty related metrics have been proposed, i.e., EPC and EPD~\citep{epc-epd}. It is important to note that these metrics are not suitable for our analysis in this paper. First, they only focus on measuring the novelty of a ranked list, while we want to not only understand the components within the predicted basket, but also analyze a model's ability to fulfill a user's needs w.r.t.\ repetition and exploration. Second, these metrics are not \ac{NBR} specific and only focus on one aspect, i.e., novelty, while we want to make a comparison between repetition and exploration to assess the \ac{NBR} performance.

To analyze the composition of a predicted basket, we propose the \emph{repetition ratio}, $\RepR$, and the \emph{exploration ratio}, $\ExplR$.
$\RepR$ and $\ExplR$ represent the proportion of \emph{repeat items} and \emph{explore items} in a recommended basket, respectively.
The overall $\RepR$ and $\ExplR$ scores are calculated over all test users as:\footnote{To assess this performance on a dataset, we use the average performance across users; we also show the corresponding user level $\RepR$ distribution in our analysis.}
\begin{equation}
\RepR = \frac{1}{N}\sum_{j=1}^{N}\frac{\left| P_{u_j}^\mathit{rep}\right|}{K}, 
\qquad
\ExplR = \frac{1}{N}\sum_{j=1}^{N}\frac{\left| P_{u_j}^\mathit{expl}\right|}{K},
\end{equation}
where $N$ denotes the number of test users, $K$ is the size of the model's predicted basket for user $u_j$, $P_{u_j}^\mathit{rep}$ and $P_{u_j}^\mathit{expl}$ are the sets of repeat items in $P_{u_j}$ and of explore items in $P_{u_j}$, respectively. 

Next, we pay attention to a basket's ability to fulfill a user's need for repetition and exploration, and propose the following metrics.
$\mathit{Recall}_\mathit{rep}$ and $\mathit{PHR}_\mathit{rep}$ are used to evaluate the $\mathit{Recall}$ and $\mathit{PHR}$ w.r.t.\ the repetition performance; similarly, we use $\mathit{Recall}_\mathit{expl}$ and $\mathit{PHR}_\mathit{expl}$ to assess the exploration performance. 
More precisely:
\begin{equation}
\mathit{Recall}_\mathit{rep} = \frac{1}{N_r}\sum_{j=1}^{N_r}\frac{\left| P_{u_j} \cap T_{u_j}^\mathit{rep}\right|}{\left| T_{u_j}^\mathit{rep}\right|}, 
\qquad
\mathit{PHR}_\mathit{rep} =  \frac{\sum_{j=1}^{N_r}\varphi\left(P_{u_j}, T_{u_j}^\mathit{rep}\right)}{N_r}
\end{equation}
and
\begin{equation}
\mathit{Recall}_\mathit{expl} \!=\! \frac{1}{N_e}\!\sum_{i=1}^{N_e}\frac{\left| P_{u_j} \cap T_{u_j}^\mathit{expl}\right|}{\left| T_{u_j}^\mathit{expl}\right|}, 
\qquad
\mathit{PHR}_\mathit{expl} \!=\! \frac{\sum_{j=1}^{N_e}\varphi\left(P_{u_j}, T_{u_j}^\mathit{expl}\right)}{N_e},
\!
\end{equation}
where $N_r$ and $N_e$ denote the number of users whose ground truth basket contains \emph{repeat items} and \emph{explore items} respectively; 
$\varphi(P, T)$ returns 1 when $P\cap T \neq \emptyset$, otherwise it returns 0.

Next, we first use the repetition ratio and exploration ratio to examine the recommended baskets; we then use our repetition and exploration metrics to re-assess the \ac{NBR} methods that we consider, examine how repetition and exploration contribute to the overall recommendation performance, and how users with different degrees of repeat behavior benefit from different \ac{NBR} methods.

\begin{figure}[t]
  \centering
  \includegraphics[clip,trim=0mm 224mm 0mm 0mm,width=0.65\linewidth]{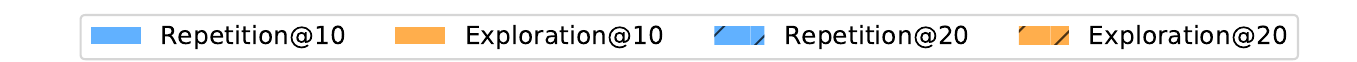}
  \\[1ex]
\begin{subfigure}{0.8\textwidth}
  \centering  
  \includegraphics[clip,trim=0mm 150mm 0mm 15mm,width=\linewidth]{figures/rep_expl_ratio1.pdf}
  \vspace*{-6mm}
  \caption{Tafeng}  
\end{subfigure}  
\begin{subfigure}{0.8\textwidth}
  \centering  
  \includegraphics[clip,trim=0mm 75mm 0mm 90mm,width=\linewidth]{figures/rep_expl_ratio1.pdf}
  \vspace*{-6mm}
  \caption{Dunnhumby}  
\end{subfigure}  
\begin{subfigure}{0.8\textwidth}
  \centering  
  \includegraphics[clip,trim=0mm 0mm 0mm 165mm,width=\linewidth]{figures/rep_expl_ratio1.pdf}
  \vspace*{-6mm}
  \caption{Instacart}  
\end{subfigure}  
  \caption{The repetition ratio $\RepR$ and exploration ratio $\ExplR$ of recommended baskets averaged over all users; solid lines indicate the repetition ratio of the ground truth; dashed lines indicate the upper bound of the repetition ratio for basket size 10 and dotted lines for basket size 20. }
  \label{fig:rep_expl_ratio}
\end{figure}

\subsection{The components of a recommended basket}
\label{section:rep-expl-components}

\begin{figure}
  \centering
\begin{subfigure}{0.8\textwidth}
    \centering
    \includegraphics[width=\linewidth]{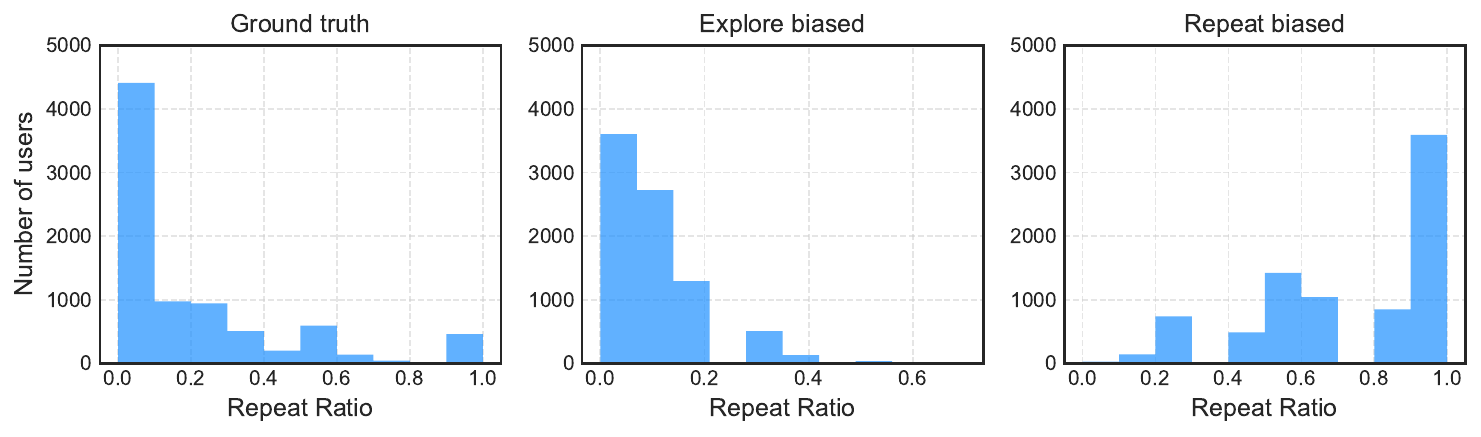}  
    \vspace*{-6mm}
    \caption{Tafeng}
    \label{fig:tafeng_repr_dist}
\end{subfigure}
\begin{subfigure}{0.8\textwidth}
    \centering
    \includegraphics[width=\linewidth]{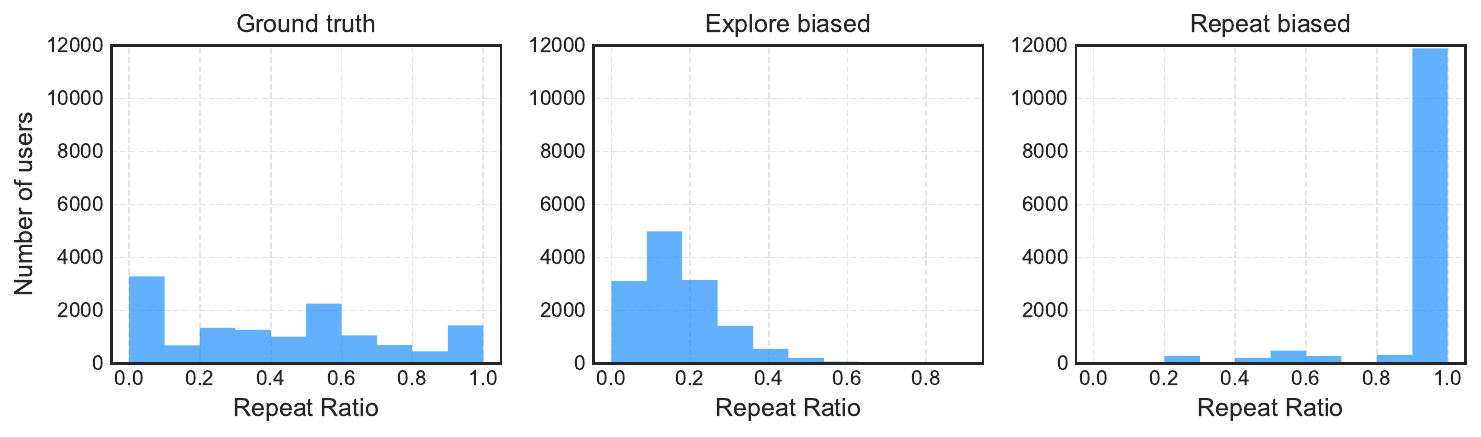}  
    \vspace*{-6mm}
    \caption{Dunnhumby}
    \label{fig:dunnhumby_repr_dist}
\end{subfigure}
\begin{subfigure}{0.8\textwidth}
  \centering
  \includegraphics[width=\linewidth]{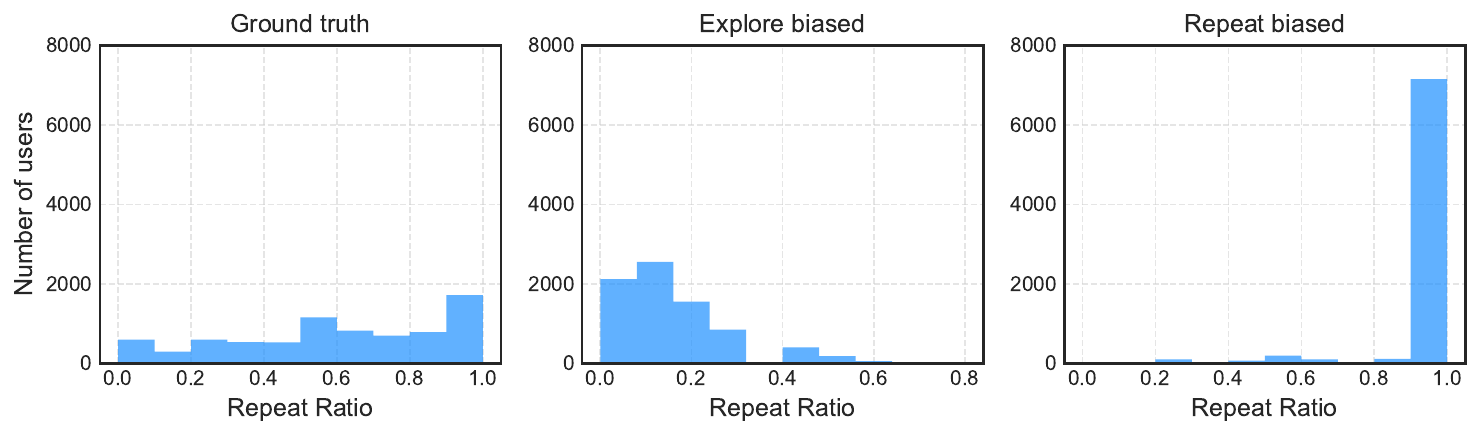}  
    \vspace*{-6mm}
  \caption{Instacart}
  \label{fig:instacart_repr_dist}
\end{subfigure}
\vspace*{2mm}
  \caption{Distribution of the repetition ratio $\RepR$ of recommended baskets on different datasets for an explore-biased method (CLEA) and a repeat-biased method (DNNTSP).}
  \label{fig:repr_distribution}
\end{figure}

We analyze the components of the recommended basket for each \ac{NBR} method to understand what makes up the recommendation.
The results are shown in Figure~\ref{fig:rep_expl_ratio}. 
First, we see that, averaged over all users, all recommended baskets are heavily skewed towards either item repetition or exploration, relative to the ground-truth baskets that are much more balanced between already seen and new items.
We can divide the methods that we compare into repeat-biased methods (i.e., P-TopFreq, GP-TopFreq, Sets2sets, DNNTSP, UP-CF@r, and TIFUKNN) and explore-biased methods (i.e., G-TopFreq, Dream, Beacon, and CLEA).
Importantly, a large performance gap exists between the two types. 
None of the published \ac{NBR} methods can properly balance the repeat items and explore items of users' future interests. 
Figure~\ref{fig:repr_distribution} shows the repetition ratio $\RepR$ distribution for the ground truth basket and the recommended basket derived by a repeat-biased method or an explore-biased method.
We show the $\RepR$ distribution of a representive explore-biased method, i.e., CLEA, and a representive repeat-biased method, i.e., DNNTSP, in Figure~\ref{fig:repr_distribution}. The $\RepR$ distribution of the other eight \acs{NBR} methods are provided in our appendix (Figure~\ref{fig:repr_distribution_app}).

\begin{table*}[!ht]
  \caption{Repetition and exploration performance comparison of frequency-based, nearest neighbor-based, and deep learning-based \ac{NBR} methods. Highlights indicate the highest score per basket size, for the \protect\hlorange{exploration} and \protect\hlblue{repetition} metrics. As in Table~\ref{tab:conventional}, we write * to indicate that the highest score for a given basket size and metric is significantly better than the second highest score (paired t-test, p-value $<0.05$).}
  \label{tab:rep-expl}
  \centering
\rotatebox{90}{%
\begin{tabular}{ll cccc cccc cccc}
    \toprule
    &Dataset & \multicolumn{4}{c}{TaFeng} & \multicolumn{4}{c}{Dunnhumby} & \multicolumn{4}{c}{Instacart}\\
    \cmidrule(r){3-6}
    \cmidrule(r){7-10}
    \cmidrule(r){11-14}
    Size & Methods & \tabincell{c}{Recall\\-rep} & \tabincell{c}{Recall\\-expl} & \tabincell{c}{PHR\\-rep}& \tabincell{c}{PHR\\-expl} &\tabincell{c}{Recall\\-rep} & \tabincell{c}{Recall\\-expl} & \tabincell{c}{PHR\\-rep}& \tabincell{c}{PHR\\-expl}&\tabincell{c}{Recall\\-rep} & \tabincell{c}{Recall\\-expl} & \tabincell{c}{PHR\\-rep}& \tabincell{c}{PHR\\-expl} \\
    \midrule
    \multirow{11}{*}{10}&
    G-TopFreq & 0.1268 & 0.0573 & 0.1947 & 0.1738 & 0.1882 & 0.0393 & 0.4954 & 0.1590 & 0.1002 & \hlorange{0.0382} & 0.4138 & \hlorange{0.1575}  \\
    &P-TopFreq& 0.5234&0.0000&0.6766&0.0000 & 0.5612 & 0.0000 & 0.8553 & 0.000 & 0.5388 & 0.0000 & 0.9085 & 0.0000\\
    &GP-TopFreq& 0.5234&0.0157&0.6766&0.0266&0.5612&0.0049&0.8553&0.0148&0.5388&0.0014&0.9085&0.0046\\
    \cmidrule{2-14}
    &UP-CF@r&\hlblue{0.6046}\Better &0.0086&\hlblue{0.7515}\Better &0.0153&\hlblue{0.5913}\Better &0.0018&\hlblue{0.8752}\Better &0.0064&0.5805&0.0011&\hlblue{0.9295}&0.0038\\
    &TIFUKNN& 0.5616 & 0.0176 & 0.7037 & 0.0284 & 0.5726 & 0.0082 & 0.8646 & 0.0257 & \hlblue{0.5931}\Better & 0.0018 & 0.9287 & 0.0060\\
    \cmidrule{2-14}
    &Dream& 0.1312 & 0.0921 & 0.1874 & 0.2378&0.1843 & 0.0412 & 0.4906 & 0.1668 & 0.1007 & 0.0367 & 0.4181 & 0.1503\\
    &Beacon&0.1456&0.0822&0.2126&0.2223&0.1893&0.0404&0.4963&0.1605&0.1010&0.0365&0.4193&0.1492\\
    &CLEA&0.1442&\hlorange{0.0935}&0.2073&\hlorange{0.2398}&0.2492&\hlorange{0.0569}\Better &0.5702&\hlorange{0.1803}\Better &0.1715&0.0337&0.5435&0.1297\\
    &Sets2Sets&0.5647 & 0.0271 & 0.7222 & 0.0495 & 0.4260 & 0.0026 & 0.7583 & 0.0064 & 0.3515 & 0.0005 & 0.7713 & 0.0013\\
    &DNNTSP& 0.5291 & 0.0556 & 0.6912 & 0.1294&0.5861 & 0.0073 & 0.8684 & 0.0224 & 0.5477 & 0.0018 & 0.9110 & 0.0054\\
    \midrule
    \multirow{11}{*}{20}
    &G-TopFreq& 0.1637 & 0.0789 & 0.2530 & 0.2385 & 0.2279 & 0.0609 & 0.5565 & 0.2353 & 0.1335 & \hlorange{0.0602} & 0.4767 & 0.2251\\
    &P-TopFreq& 0.7251 & 0.0000& 0.8439&0.0000&0.7399 & 0.0000 & 0.9353 & 0.0000& 0.7193 & 0.0000 & 0.9631 & 0.0000\\
    &GP-TopFreq& 0.7251 & 0.0339& 0.8439&0.0724&0.7399&0.0153&0.9353&0.0470&0.7193&0.0084&0.9631&0.0285\\
    \cmidrule{2-14}
    &UP-CF@r&\hlblue{0.7785}\Better&0.0259&\hlblue{0.8742}&0.0560&0.7718&0.0078&\hlblue{0.9430}\Better &0.0261&0.7612&0.0083&\hlblue{0.9735}&0.0215\\
    &TIFUKNN& 0.7664 & 0.0410 & 0.8707 &0.0806 &0.7474 & 0.0159 & 0.9344 & 0.0677 & \hlblue{0.7747}\Better & 0.0108 & 0.9728 & 0.0318\\
    \cmidrule{2-14}
    &Dream& 0.1723 & \hlorange{0.1234} & 0.2572 & \hlorange{0.3264} & 0.2316 & 0.0687 & 0.5612 & 0.2596& 0.1348 & 0.0586 & 0.4849 & 0.2222\\
    &Beacon&0.1748&0.1230&0.2621&0.3262&0.2333&0.0672&0.5628&0.2568&0.1338&0.0599&0.4822&\hlorange{0.2252}\\
    &CLEA&0.1799&0.1217&0.2675&0.3189&0.2808&\hlorange{0.0874}\Better &0.6168&\hlorange{0.2648}\Better &0.2081&0.0578&0.6051&0.2044\\
    &Sets2Sets& 0.7478 & 0.0558 & 0.8703 & 0.1166&0.6317 & 0.0086 & 0.8881 & 0.0273 &0.5095 & 0.0032 & 0.8873 & 0.0073\\  
    &DNNTSP& 0.6820 & 0.0899 & 0.8130 & 0.2103 &\hlblue{0.7758} & 0.0183 & 0.9391 & 0.0609&0.7284 & 0.0095 & 0.9657 & 0.0320\\
    \bottomrule
  \end{tabular}}
\end{table*}

Among the explore-biased methods, G-TopFreq is not a personalized method; it always provides the most popular items. 
Dream, Beacon, and CLEA treat all items without any discrimination, which means the explore items are more likely to be in the predicted basket and their basket components are similar to G-TopFreq. 
Looking at the performance in Table~\ref{tab:conventional}, we see that repeat-biased methods generally perform much better than explore-biased methods on conventional metrics across the datasets, especially when the dataset has a relatively high repetition ratio.

The repetition ratios of P-TopFreq and GP-TopFreq serve as the upper bound repetition ratio for the recommended basket. Most baskets recommended by repeat-biased methods are close to or reach this upper bound, even when the datasets have a low ratio of repeat behavior in the ground truth, except for two cases (Sets2sets and DNNTSP on the TaFeng dataset). 

Finally, the exploration ratio of repeat-biased methods increases from basket size 10 to 20; we believe that this is because there are simply no extra repeat items available: it does not mean that the methods actively increase the exploration ratio in a larger basket setting.

\subsection{Performance w.r.t.\ repetition and exploration}
The results in terms of repetition and exploration performance are shown in Table~\ref{tab:rep-expl}. 
First of all, using our proposed metrics, we observe that the repetition performance $\mathit{Recall}_\mathit{rep}$ is always higher than the exploration performance, even when the explore items form almost 90\% of the recommended basket. This shows that the repetition task (recommending repeat items) and the exploration task (recommending explore items) have different levels of difficulty and that capturing users' repeat behavior is much easier than capturing their explore behavior. 

Three deep learning-based methods perform worst w.r.t.\  repeat item recommendation \emph{and} best w.r.t.\  explore item recommendation at the same time, as they are heavily skewed towards explore items. 
We also see that there are improvements in the exploration performance compared to G-TopFreq with the same level of exploration ratio, which indicates that the representation learned by these methods does capture the hidden sequential transition relationship between items. Repeat-biased methods perform better w.r.t.\ repetition in all settings, since the baskets they predict contain more repeat items. Similarly, we can see that DNNTSP, UP-CF@r, and TIFUKNN perform better than P-TopFreq w.r.t.\  repeat performance with the same or a lower level of repetition ratio.

Third, explore-biased methods spend more resources on the more difficult and uncertain task of explore item prediction, which is not an optimal choice when considering the overall \ac{NBR} performance. 
Being biased towards the easier task of repeat item prediction leads to gains in the overall performance, which is positively correlated with the repetition ratio of the dataset. 

To understand the potential reasons for a method being repeat-biased or explore-biased, we provide an in-depth analysis of the methods' architectures. P-TopFreq and GP-TopFreq are repeat-biased methods as they both mainly rely on the frequency of historical items to recommend the next basket. 
Two nearest neighbor-based methods, i.e., TIFUKNN and UP-CF@r, have a module to model both the frequency and the recency of historical items; besides, they both have a parameter to emphasize the frequency and recency information. Similarly, Sets2sets is also repeat-biased as it adds the historical items' frequency information to the prediction layer. 
DNNTSP does not consider frequency information, however, it has an indicator vector to indicate whether an item has appeared in the historical basket sequence or not, which can be regarded as a repeat item indicator.
G-TopFreq is explore-biased since it is not a personalized method and can only recommend top-$k$ popular items within the dataset. The remaining three explore-biased methods (Dream, Beacon, and CLEA) do not consider the frequency of historical items or the indicator of items' appearance, so they fail to identify the benefits of recommending repeat items.

\subsection{The relative contribution of repetition and exploration}

Even though a clear improvement w.r.t.\ either repeat or explore performance can be observed in the previous section, this does not mean that this improvement is the reason for the better overall performance, since repeat and explore items account for different ground truth proportions in different datasets. 
To better understand where the performance gains of the well-performing methods in Table~\ref{tab:conventional} come from, we remove \emph{explore items} and keep \emph{repeat items} in the predicted basket to compute the contribution of repetition, similarly, we remove \emph{repeat items} and keep \emph{explore items} to compute the performance, which can be regarded as the contribution of exploration.

\begin{figure}[t]
    \centering
  \includegraphics[clip,trim=0mm 224mm 0mm 0mm,width=0.65\linewidth]{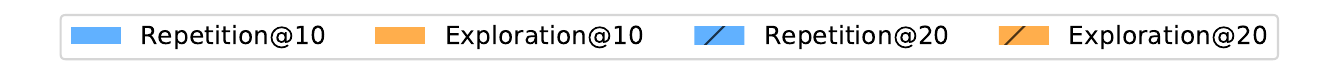}
  \\[1ex]
\begin{subfigure}{0.8\textwidth}
  \centering  
  \includegraphics[clip,trim=0mm 150mm 0mm 15mm,width=\linewidth]{figures/re_performance_gain.pdf}
  \vspace*{-6mm}
  \caption{Tafeng}  
\end{subfigure}  
\begin{subfigure}{0.8\textwidth}
  \centering  
  \includegraphics[clip,trim=0mm 75mm 0mm 90mm,width=\linewidth]{figures/re_performance_gain.pdf}
  \vspace*{-6mm}
  \caption{Dunnhumby}  
\end{subfigure}  
\begin{subfigure}{0.8\textwidth}
  \centering  
  \includegraphics[clip,trim=0mm 0mm 0mm 165mm,width=\linewidth]{figures/re_performance_gain.pdf}
  \vspace*{-6mm}
  \caption{Instacart}  
\end{subfigure}      
    \caption{Performance contribution from repeat and explore recommendations on the Tafeng, Dunnhumby, and Instacart datasets.}
    \label{fig:performance_contribute}
\end{figure}

Experimental results on three datasets are shown in Figure~\ref{fig:performance_contribute}.
We consider G-TopFreq, P-TopFreq, and GP-TopFreq as simple baselines to compare with.
From Figure~\ref{fig:performance_contribute}, we conclude that Dream and Beacon perform better than G-TopFreq on the TaFeng dataset, as the main performance gain is from improvements in the exploration prediction. 
As a consequence, in the Dunnhumby and Instacart datasets, Dream, Beacon, and G-TopFreq achieve similar performance, and the repeat prediction contributes the most to the overall performance, even  when their recommended items are heavily skewed towards explore items. Also, we observe that CLEA outperforms other explore-biased methods due to its improvements in the repetition performance without sacrificing the exploration performance. 

At the same time, it is clear that TIFUKNN, UP-CF@r, Sets2Sets, and DNNTSP outperform explore-biased methods because of the improvements in the repetition performance, even at the detriment of exploration. 
The repeat items make up the majority of their correct recommendations. 
Specifically, repeat recommendations contribute to over 97\% of their overall performance on the Dunnhumby and Instacart datasets.

An interesting comparison is between Sets2Sets and P-TopFreq. The strong performance gain of Sets2sets on the TaFeng dataset is mainly due to the exploration part, whereas P-TopFreq outperforms it by a large margin on the other two datasets at the same level of repetition ratio, even though the personal frequency information is considered in the Sets2sets model. 
We believe this indicates that the loss on repeat items seems to be suppressed by the loss on explore items during the training process, which weakens the influence of the frequency information.

Recall that the number of repetition candidates for a user may be smaller than the basket size, which means that there might be empty slots in the basket recommended by P-TopFreq. 
From Figure~\ref{fig:rep_expl_ratio} and Table~\ref{dataset: rep_slots}, we observe that the empty slots account for a significant proportion of exploration slots in many settings. 
However, existing studies omit this fact when making the comparison with P-TopFreq, leading to an unfair comparison and overestimation of the improvement, as their predictions leverage more slots. For example, Dream, Beacon, and CLEA can beat P-TopFreq, but they are inferior to GP-TopFreq. TIFUKNN and UP-CF@r model the temporal order of the frequency information, leading to a higher repetition performance than P-TopFreq in general. Even though the contribution of the repetition performance improvement is obvious on the Instacart dataset, it is less meaningful on the other two datasets, where the performance gain is mainly from the exploration part by filling the empty slots. 
When compared with the proposed GP-TopFreq baseline on the Tafeng and Dunnhumby datasets, the improvement is around a modest 3\%. 

DNNTSP is always among the best-performing methods across the three datasets and is able to model exploration more effectively than other repeat-biased methods. 
Moreover, it also actively recommends explore items, rather than being totally biased towards the repeat recommendation in high exploration scenarios. 
However, the improvement is limited due to the relatively high repetition ratios and the huge difficulty gap between repetition and exploration tasks. 
Compared with GP-TopFreq, the improvement of DNNTSP w.r.t.\ $\mathit{Recall}@10$ on the Dunnhumby and Instacart datasets is merely 1.3\% and 1.9\% respectively, which is modest considering the complexity and training time added by DNNTSP. 

Obviously, even though many advanced \ac{NBR} algorithms learn rich user and/or item representations, the main performance gains stem from the prediction of repeat behavior. 
Yet, limited progress w.r.t. overall performance has so far been made compared to the simple P-TopFreq and GP-TopFreq baseline methods.

\begin{figure*}[t]
  \begin{subfigure}{\textwidth}
  \centering
  \includegraphics[width=\textwidth]{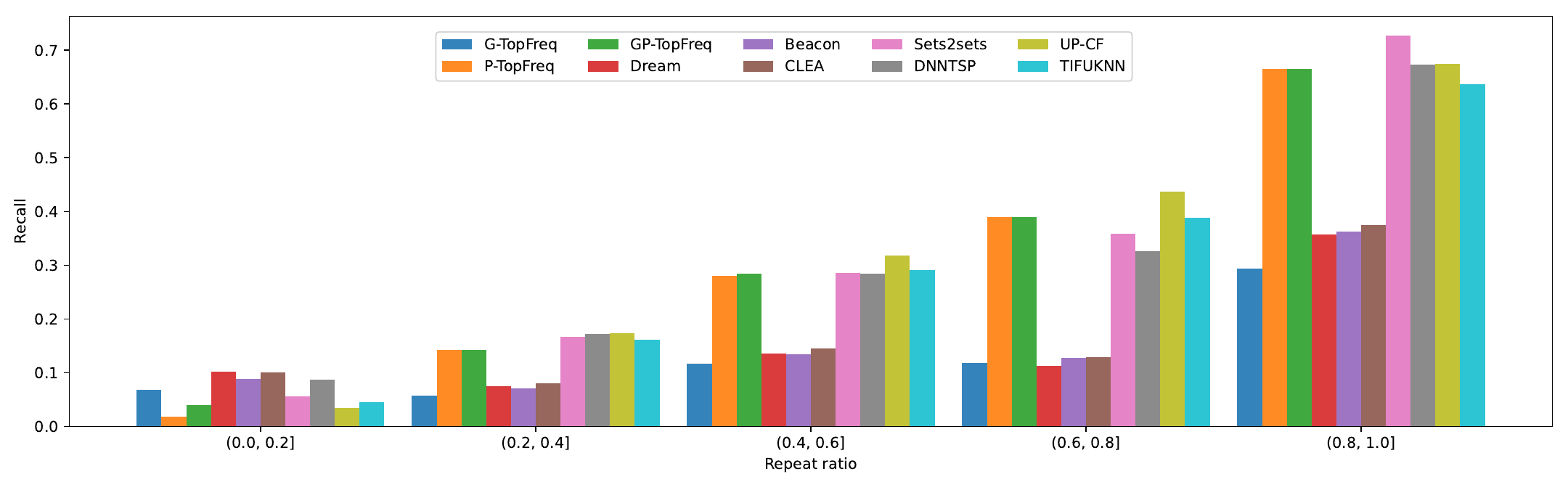}
  \caption{Tafeng}
  \end{subfigure}
  
  \vspace*{2mm}
  \begin{subfigure}{\textwidth}
  \centering
  \includegraphics[width=\textwidth]{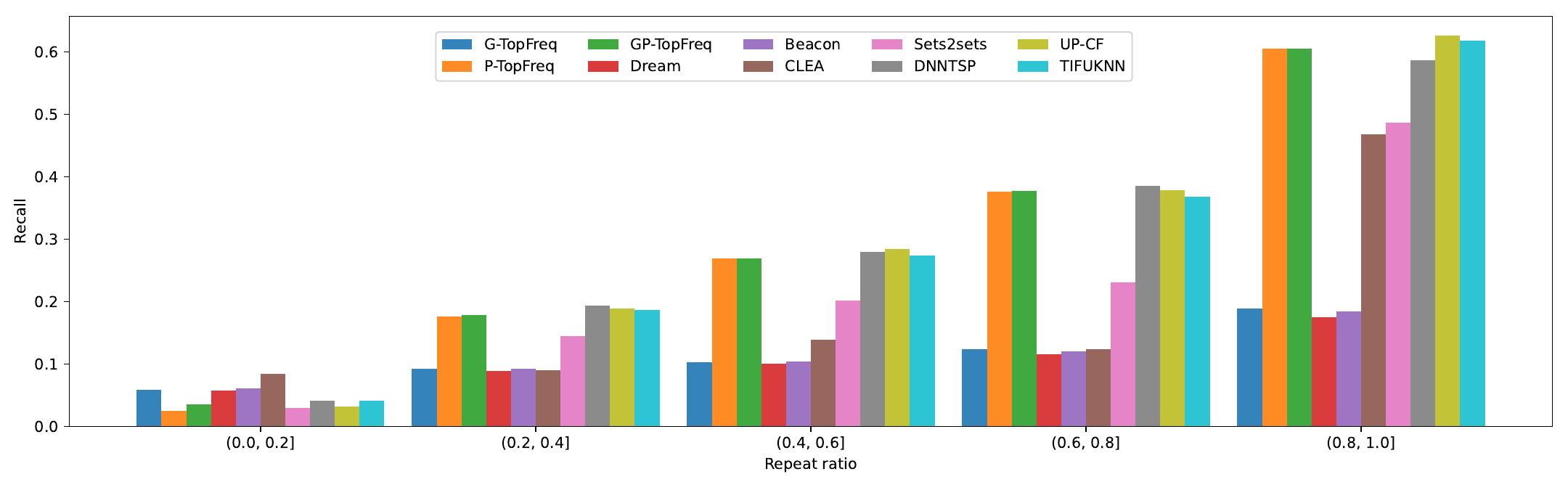}
  \caption{Dunnhumby}
  \end{subfigure}
  
  \vspace*{2mm}
  \begin{subfigure}{\textwidth}
  \centering
  \includegraphics[width=\textwidth]{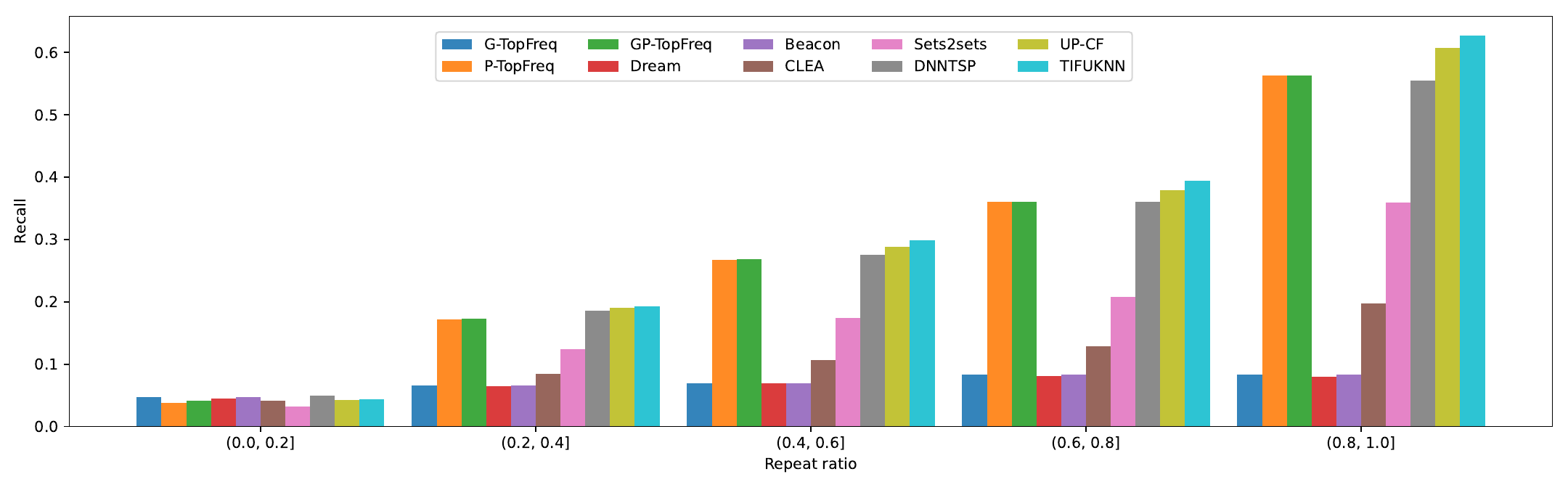}
  \caption{Instacart}
  \end{subfigure}  
  \caption{Treatment effect for users on the Tafeng, Dunnhumby, and Instacart datasets, for ten \ac{NBR} methods for users with different repetition ratios (binned in five groups).}
  \label{fig:user_level_analysis_performance}
\end{figure*}

\subsection{Treatment effect for users with different repetition ratios}
\label{sec: treatment-users}

As the average repetition ratio in a dataset has a significant influence on a model's performance (see Section~\ref{section:overall}), existing \acs{NBR} methods are skewed to repetition or exploration (see Section~\ref{section:rep-expl-components}) and global trend might influence the users repetition patterns, it is of interest to investigate the treatment effect for users with different repetition ratios. 
We examine the performance of \ac{NBR} methods w.r.t.\ different groups of users with different repetition ratios. 
We divide the users into 5 groups according to their repetition ratio: $[0, 0.2]$, $(0.2, 0.4]$, $(0.4, 0.6]$, $(0.6, 0.8]$, $(0.8, 1.0]$, and calculate the average performance within each group. 
Note that the repetition ratio indicates the user's preference w.r.t.\ \emph{repeat items} and \emph{explore items}, e.g., users with a low repetition ratio prefer to purchase new items in their next basket.
The results are shown in Figure~\ref{fig:user_level_analysis_performance}.

First, we can see that the methods' performance within different user groups is different from the performance computed over all users (Table~\ref{tab:conventional}). For example, several explore-biased methods (G-TopFreq, Dream, Beacon, CLEA) can outperform recent repeat-biased methods (TIFUKNN, UP-CF@r, Sets2Sets, DNNTSP) in the user group with a low repetition ratio, $[0, 0.2]$, but these explore-biased methods are inferior to the repeat-biased methods when computing the performance over all users.
Second, the performance of repeat-biased \acs{NBR} models increases, as the repetition ratio increases. 
Interestingly, we observe an analogous trend w.r.t.\ the performance of explore-biased \acs{NBR} methods as the repetition ratio increases, but the rate of the increase is smaller. 
We believe that this is because the \acs{NBR} task gets easier for users with a higher repetition ratio, and the repeat-biased methods benefit more from an increase in repetition ratio. 

From the perspective of user group fairness, explore-biased methods seem to be fairer than repeat-biased methods across different user groups, as they have a very similar performance across groups.
Explore-biased methods have lower variation in performance than repeat-biased methods. However, we should be aware of intrinsic difficulty gaps between different user groups, e.g., it is easier for \acs{NBR} methods to find correct items for users who like to repeat purchase. 
Taking this into consideration, we take G-TopFreq and GP-TopFreq as two anchor baselines to evaluate whether recent \acs{NBR} methods put a specific user group at a disadvantage or not. On the Tafeng and Dunnhumby datasets, repeat-biased methods (Sets2Sets, UP-CF, TIFUKNN, DNNTSP) fail to achieve the performance of G-TopFreq within users whose repetition ratio is in $[0, 0.2]$, which means they do not cater to users of this group.
At the same time, recent explore-biased methods (Dream, Beacon, CLEA) fail to achieve the performance derived by the very simple baseline, i.e., GP-TopFreq, on four user groups on Tafeng, Dunnhumby, and Instacart dataset.
This analysis indicates that both repeat-biased and explore-biased \acs{NBR} methods do not treat all user groups fairly.

\subsection{Looking beyond the average performance}
\label{sec: average-performance}

In the recommender systems literature it is customary to compute the average performance over all test users to represent the performance of a recommendation method.
Given the diverse treatment effect across different user groups, we want to drill down and see how much the different user groups contribute towards the overall average performance.
As before, we use five groups as defined in Section~\ref{sec: treatment-users} in terms of the repetition ratio.  Specifically, for each individual group $g_j$, we analyze its \emph{proportion of all users} ($\mathit{PAU}$) and its \emph{contribution to the average performance} ($\mathit{CAP}$) as follows:
\begin{align}
  \mathit{PAU}_j &= \frac{|U_{g_j}|}{\sum_{j=1}^{q}{|U_{g_j}|}}
\\
  \mathit{CAP}_j &= \frac{\sum_{u\in{U_{g_j}}}{\mathit{Perf}_u}}{\sum_{j=1}^{q}{\sum_{u\in{U_{g_j}}}{\mathit{Perf}_u}}},
\end{align}
where $U_{g_j}$ denotes the set of users in group $g_j$, $q$ denotes the number of user groups, $\mathit{Perf}_u$ represents the method's performance w.r.t.\ user $u$. Note that the performance metric we analyze in this section is $Recall@10$, but similar phenomena can be observed for other metrics.

The results in terms of $\mathit{PAU}$ and $\mathit{CAP}$ are shown in Table~\ref{tab: average_metrics_contribution}. Under the ideal circumstances, the contribution to the average performance $\mathit{CAP}$ of each user group should be equal to its proportion of all users $\mathit{PAU}$; this would allow us to use the average performance of a method as its overall performance and leave no user group behind.
However, we can see that $\mathit{CAP}_{(0.8, 1]}$ is much higher than $\mathit{PAU}_{(0.8, 1]}$ and $\mathit{CAP}_{[0, 0.2]}$ is much lower than $PAU_{[0, 0.2]}$ for every \acs{NBR} method (both repeat-biased methods and explore-biased methods) on all datasets. On the Tafeng dataset, only 5.5\% of the users belong to group $(0.8, 1]$. However, their contribution to the average performance ranges from 18.8\% to 36.8\%. On the Dunnhumby dataset, 31.4\% of the users belong to group $[0, 0.2]$, while the $\mathit{CAP}_{[0, 0.2]}$ for repeat-biased methods (i.e., P-TopFreq, GP-TopFreq, Sets2Sets, DNNTSP, TIFUKNN and UP-CF@r) only ranges from 3.1\% to 5.0\%.
The results reflect that there might be a long-tail distribution w.r.t.\ the user's contribution to the average performance (i.e., few users contribute a large proportion to the performance), since the \acs{NBR} task for different users might have different difficulty levels.

Given the previous observations, we construct a simple example to demonstrate the potential limitations of average performance. Assume we have two user groups, i.e., group $g_a$ with 10 users and group $g_b$ with only 1 user, where the \ac{NBR} task for $g_b$ is easier than $g_a$. Assume, also, that we have a baseline method $M_b$ whose performance $\mathit{Perf}$ can achieve 0.02 in group $g_a$ and 0.4 in group $g_b$. We have another two optimized methods $M_{\alpha}$ and $M_{\beta}$. Compared to baseline $M_b$, $M_{\alpha}$ can achieve 100\% improvement in group $g_a$, $M_{\beta}$ can also achieve 100\% improvement in group $g_b$ at the cost of 50\% reduction in group $g_a$.
In this case, the cumulative improvement of $M_{\alpha}$ is $0.02 \times 10=0.2$, while the cumulative improvement of $M_{\beta}$ is $0.4 \times 1 - 0.01 \times 10 = 0.3$. $M_{\beta}$ is considered to be better than $M_{\alpha}$, since $M_{\beta}$'s average performance is higher. However, we notice that $M_{\alpha}$ can improve the performance of 10 users, while $M_{\beta}$ can only improve the performance of 1 user and at the detriment of the other 10 users.
To sum up, the average performance has limitations to represent the performance of methods on the \acs{NBR} task and it might put users in a specific group at disadvantage.\footnote{In this paper, to remain focused we only analyze the repetition-exploration issue. However, there might be other factors (e.g., basket size, historical basket length) that also influence the difficulty level of \ac{NBR} problem.} We should calculate the performance of each user group in order to have a comprehensive understanding of the \acs{NBR} method.
\begin{table}[ht]
	\setlength\tabcolsep{0.12cm}
	\centering
	\caption{Group proportion of all users ($\mathit{PAU}$) and contribution to the average performance ($\mathit{CAP}$).}
  \label{tab: average_metrics_contribution}
	\begin{tabular}{ l l  c c c c c }
		\toprule
		 &  & \multicolumn{5}{c}{User group}\\
		\cmidrule{3-7}
		Dataset & Method & [0.0, 0.2]& (0.2, 0.4] & (0.4, 0.6] & (0.6, 0.8] & (0.8, 1.0] \\
		\midrule
		
		 \multirow{11}{*}{TaFeng} 
    & $\mathit{PAU}$ & 68.8\%&14.9\%&8.5\%&2.4\%&5.5\%\\
    \cmidrule{2-7}
    & G-TopFreq &53.4\%&10.2\%&10.4\%&3.3\%&22.7\%\\
    & P-TopFreq &12.3\%&20.9\%&21.9\%&8.2\%&36.8\%\\
    & GP-TopFreq &21.9\%&18.8\%&19.5\%&7.3\%&32.5\%\\
    & Dream &60.3\%&9.7\%&9.0\%&2.2\%&18.8\%\\
    & Beacon &59.0\%&9.3\%&9.4\%&2.5\%&19.9\%\\
    & CLEA &60.4\%&9.3\%&9.2\%&2.3\%&18.9\%\\
    & Sets2Sets &27.9\%&18.2\%&18.3\%&6.4\%&29.1\%\\
    & DNNTSP&37.4\% &16.3\%&16.0\% &5.5\% &24.7\%\\
    & TIFUKNN &24.2\%&19.8\% & 19.8\%&7.1\% &29.0\% \\
		& UP-CF@r &18.4\%& 21.2\%&21.4\%&8.4\%&30.8\%\\
		\midrule
		\multirow{11}{*}{Dunnhumby}
    & $\mathit{PAU}$ & 31.4\%&19.6\%&21.8\%&14.3\%&12.9\%\\
    \cmidrule{2-7}
    & G-TopFreq &16.5\%&18.6\%&25.2\%&18.1\%&21.6\%\\
    & P-TopFreq &3.1\%&14.8\%&25.7\%&22.6\%&33.8\%\\
    & GP-TopFreq &4.3\%&14.7\%&25.5\%&22.3\%&33.3\%\\
    & Dream &16.8\%&19.2\%&24.9\%&17.9\%&21.2\%\\
    & Beacon &16.8\%&18.7\%&24.8\%&17.9\%&21.8\%\\
    & CLEA &15.4\%&12.2\%&20.0\%&12.2\%&40.1\%\\
    & Sets2Sets &4.2\%&16.1\%&25.1\%&19.7\%&34.9\%\\
    & DNNTSP &5.0\%&15.7\%&25.6\%&22.5\%&31.2\%\\
    & TIFUKNN &4.9\%&15.1\%&25.1\%&22.0\%&32.9\%\\
    & UP-CF@r &3.7\%&15.3\%&25.2\%&22.3\%&33.5\%\\
    \midrule
    \multirow{11}{*}{Instacart}
    & $\mathit{PAU}$ & 13.2\%&14.9\%&20.0\%&21.9\%&30.0\%\\
    \cmidrule{2-7}
    & G-TopFreq & 8.9\%&12.3\%&19.7\%&24.0\%&35.0\%\\
    & P-TopFreq &1.6\%&7.8\%&16.4\%&23.7\%&50.4\%\\
    & GP-TopFreq &1.8\%&7.9\%&16.4\%&23.6\%&50.3\%\\
    & Dream &9.2\%&12.3\%&19.8\%&23.8\%&34.8\%\\
    & Beacon &8.9\%&12.4\%&19.5\%&23.5\%&35.6\%\\
    & CLEA &7.0\%&8.6\%&14.1\%&17.6\%&52.6\%\\
    & Sets2Sets &2.0\%&8.7\%&15.8\%&21.3\%&52.2\%\\
    & DNNTSP &2.0\%&8.1\%&16.5\%&23.3\%&50.1\%\\
    & TIFUKNN &1.8\%&8.0\%&16.3\%&23.6\%&50.3\%\\
    & UP-CF@r &1.8\%&8.0\%&16.5\%&23.5\%&50.2\%\\
		\bottomrule			
	\end{tabular}	
\end{table}

\subsection{Treatment effect for items with different frequencies}
\begin{figure*}
  \begin{subfigure}{\textwidth}
  \centering
  \includegraphics[width=\textwidth]{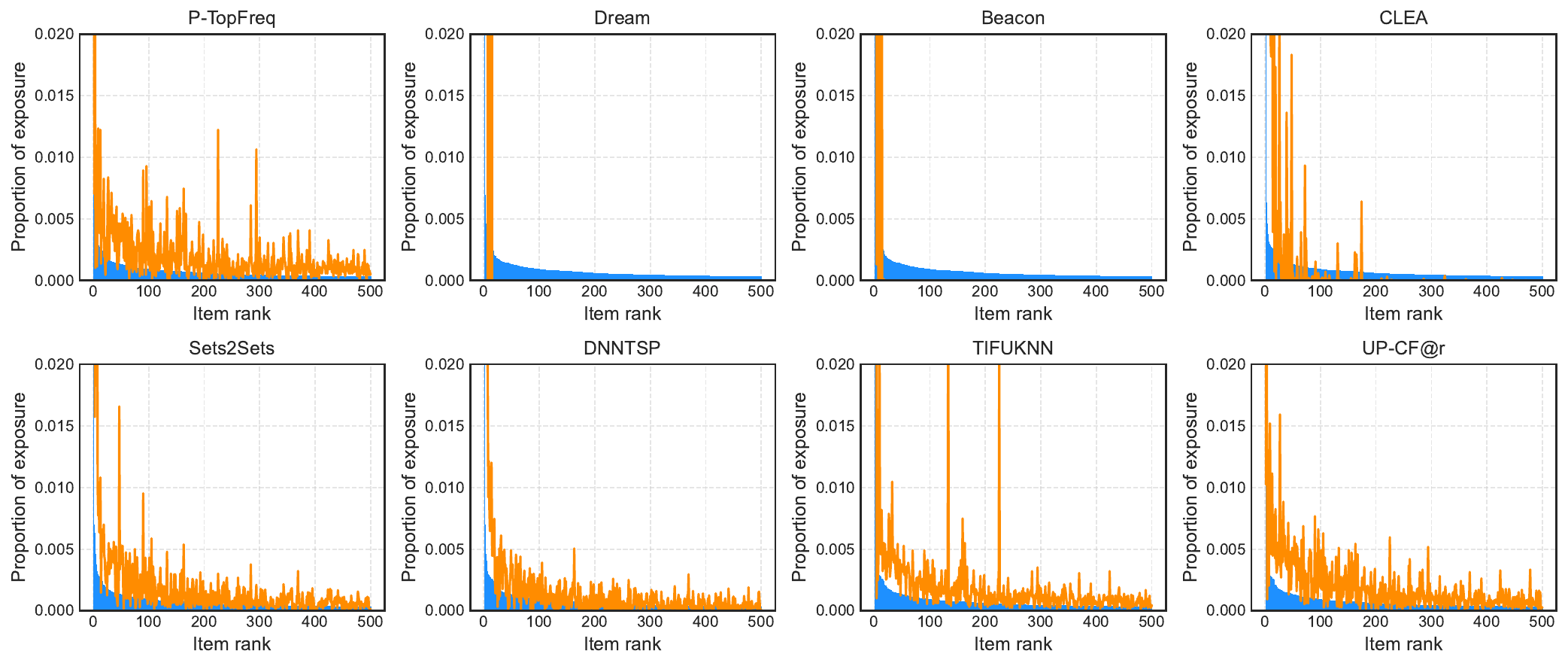}
  \caption{Items ranked according to their frequency in the training labels.}
  \label{fig:exposure_labels}
  \end{subfigure}
  
  \vspace*{2mm}
  \begin{subfigure}{\textwidth}
  \centering
  \includegraphics[width=\textwidth]{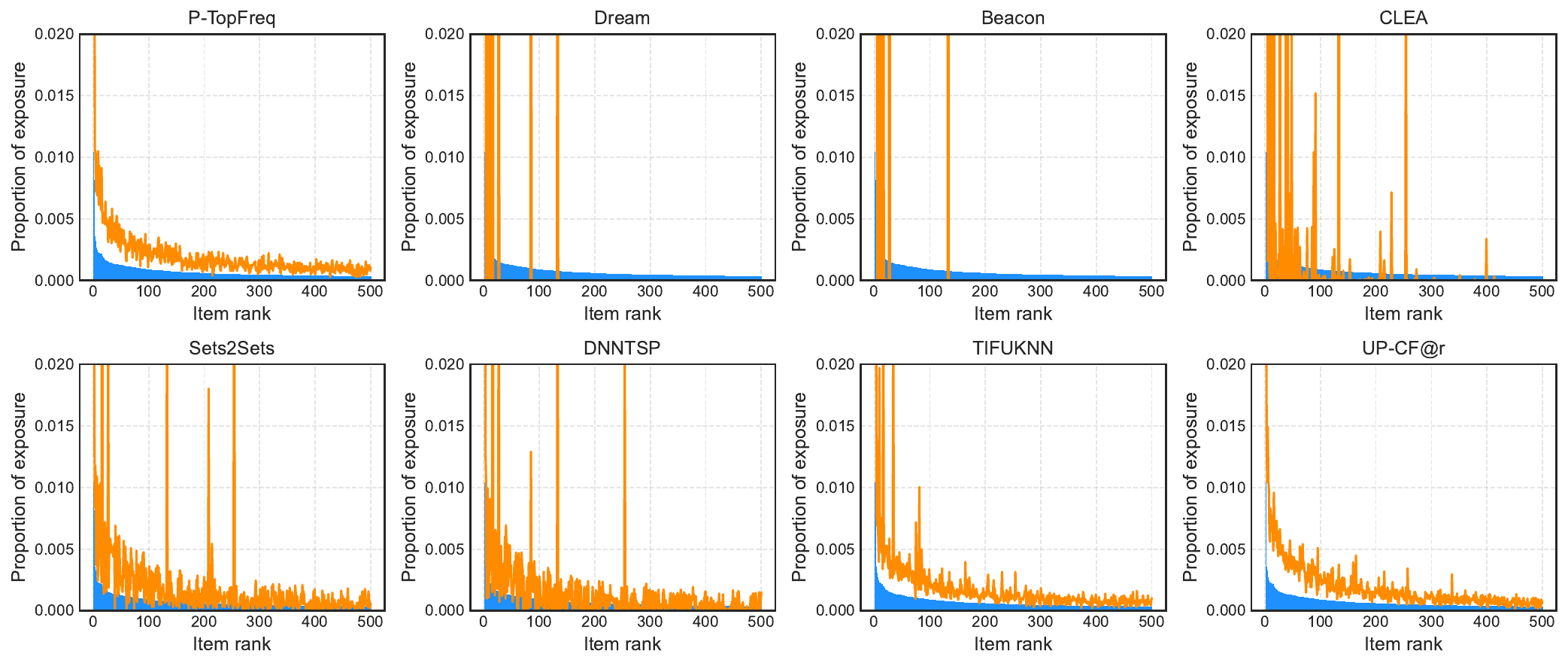}
  \caption{Items ranked according to their frequency in the test inputs (i.e., historical baskets of test users).}
  \label{fig:exposure_historical}
  \end{subfigure}
  \vspace*{0mm}
  \caption{Treatment effect for items on the Tafeng dataset, for eight \ac{NBR} methods for items with different frequencies in training labels and testing inputs. The blue bar shows the frequency distribution, and the orange line denotes the exposure distribution.}
  \label{fig:item_level_exposure_analysis}
\end{figure*}

The \Ac{NBR} scenario can be thought of in terms of a two-sided market with items and users~\citep{wang2021user,patro2020fairrec,biswas2021toward}. So far, we have analyzed the user-side performance from several aspects. 
In this section, we analyze treatment effects of \ac{NBR} methods from the item side. Specifically, we investigate the relation between an item's exposure and its frequency in training labels (the ground-truth items of the training users) or test inputs (the historical items of the test users). As the item exposure in recommended baskets and the item frequency have different scales, we use the exposure and frequency of all items, respectively, to normalize them.
In order to visualize the relation between an item's exposure and its frequency, we rank items according to their frequency and select the top-500 items. The frequency and the exposure distributions for different methods on the Tafeng dataset are shown in Figure~\ref{fig:item_level_exposure_analysis}.\footnote{Experimental results on the Dunnhumby and Instacart datasets are provided in the appendix, and qualitatively similar patterns can be observed. We do not include G-TopFreq in this analysis, since it always recommends top-$K$ items in the historical dataset.}

First, we observe the long-tail distribution w.r.t.\ the item exposure in all \acs{NBR} methods; a small number of items get a large proportion of the total exposure. 
Surprisingly, a large proportion of items do not get any exposure in the baskets recommended by Dream, Beacon and CLEA on the TaFeng dataset, which we consider to be sub-optimal from the item perspective.
Second, the item exposure distributions of P-TopFreq, TIFUKNN, and UP-CF@r are more related to the frequency distribution of the test input than to the  frequency distribution of the training labels.  We believe that the repeat-biased nature of those algorithms, as well as the absence of training, results in recommendations that are strongly dependent on the items' frequency in historical baskets, i.e., on the test inputs. 
Third, in deep learning-based methods (Dream, Beacon, CLEA, Sets2Sets, and DNNTSP), we can see that the distribution of items with high exposure shifts to the left, from Figure~\ref{fig:exposure_historical} to Figure~\ref{fig:exposure_labels}. This result reflects the fact that an item's high exposure is more closely related to its high frequency in the training labels.
To sum up, the item frequency distributions in the training labels and test inputs have a different impact on the item exposure of different \acs{NBR} methods.

\subsection{Upshot}
\label{sec: upshot-rep-expl}
Based on our second round of analyses of state-of-the-art \ac{NBR} methods that we conducted with purpose-built metrics, we observe that there is a clear difficulty gap and trade-off between the repetition task and the exploration task. 
As a rule of thumb, being biased towards the easier repetition task is an important strategy that helps to boost the overall \ac{NBR} performance. 
Deep learning-based methods do not effectively exploit the repetition behavior. 
Indeed, they achieve a relatively good exploration performance, but they are not able to outperform the simple frequency baseline GP-TopFreq in several cases. 
Some recent state-of-the-art \ac{NBR} methods are skewed towards the repetition task and outperform GP-TopFreq.
However, the improvements they achieve are limited, especially considering the complexity and computational costs, e.g., for the training process~\cite{dnntsp} and for hyper-parameters search~\cite{recency, tifuknn}.

Moreover, current \acs{NBR} methods usually focus on improving the overall performance, but they often fail to provide, or exploit, deeper insights into the components of their recommended baskets (skewed towards repetition or exploration).

Furthermore, different \acs{NBR} methods have different treatment effects across different user groups, and the widely-used average performance can not fully evaluate the models' performance, e.g., methods might achieve high overall performance at the detriment of a specific user group, which accounts for a large proportion of all users.
From the item-side perspective, few items account for a large proportion of the total exposure in all \acs{NBR} methods, and some \acs{NBR} methods might only recommend a small set of items to users.

\section{Conclusion}
We have re-examined the reported performance gains of deep learning-based methods for the \acf{NBR} task over frequency-based and nearest neighbor-based methods.
We analyzed state-of-the-art \acs{NBR} methods on the following seven aspects:
\begin{enumerate*}[label=(\roman*)]
\item the overall performance on different scenarios;
\item the basket components;
\item the repeat and explore performance; 
\item the contribution of repetition and exploration to the overall performance; 
\item the treatment effect for different user groups;
\item the potential limitations of the average metrics; and
\item the treatment effect for different items.
\end{enumerate*}

\subsection{Main findings}
We arrive at several important findings:
\begin{enumerate*}[label=(\roman*)]
\item No state-of-the-art \ac{NBR} method, deep learning-based or otherwise, consistently shows the best performance across datasets. Compared to a simple frequency-based baseline, the improvements are modest or even absent.
\item There is a clear difficulty gap and trade-off between the repeat task and the explore task. 
As a rule of thumb, being biased towards the easier repeat task is an important strategy that helps to boost the overall \ac{NBR} performance.
\item Some \acs{NBR} methods might achieve better average overall performance at the detriment of a user group with a large proportion of users.
\item Deep learning-based methods do not effectively exploit repeat behavior. They indeed achieve relatively good explore performance, but are not able to outperform the simple frequency-based baseline GP-TopFreq in terms of the relatively easy repetition task. 
Some state-of-the-art \ac{NBR} methods are skewed towards the repeat task and because of this they are able to outperform GP-TopFreq; however, their improvements are limited, especially considering their added complexity and computational costs.
\end{enumerate*}

\subsection{Insights for \acs{NBR} model evaluation}
Our work highlights the following important guidelines that practitioners and researchers working on \ac{NBR} should follow when evaluating or designing an \ac{NBR} model:
\begin{enumerate*}[label=(\roman*)]
\item Use a diverse set of datasets for evaluation, with different ratios of repeat items and explore items; 
\item Use GP-TopFreq as a baseline when evaluating \ac{NBR} methods;
\item Apart from the conventional accuracy-based metrics, consider the newly introduced repeat and explore metrics, $\mathit{Recall}_\mathit{rep}$, $\mathit{PHR}_\mathit{rep}$, $\mathit{Recall}_\mathit{expl}$ and $\mathit{PHR}_\mathit{expl}$, as a set of fundamental metrics to understand the performance of \ac{NBR} methods; 
\item The $\mathit{RepR}$ and $\mathit{ExplR}$ statistics should be included to understand what kind of items shape the recommended baskets; and
\item Calculate the performance of each user group to get a comprehensive understanding of the \acs{NBR} methods.
\end{enumerate*}

\subsection{Insights for \acs{NBR} model design}
From the analysis of this paper, apart from the difficulty imbalance between the repetition and exploration task, we should also be aware that the repetition recommendation task and exploration recommendation task have different characteristics.
For instance, the repetition recommendation task focuses on predicting whether historical items will be repurchased or not, where the frequency and recency of historical items are very important, and the exploration recommendation task focuses on inferring \emph{explore items} from a much bigger search space via modeling item-to-item correlations, which deep-learning methods might be good at.
Therefore, just blindly designing complex \ac{NBR} models without considering the difference between repetition and exploration might be sub-optimal.

We think that it is better to separate the repetition recommendation and exploration recommendation in the \acs{NBR} task (e.g. using frequency and recency to address the repetition task, and using NN-based models to model item-to-item correlations), which not only allows us to address repetition and exploration according to their respective characteristics but also offers the flexibility of controlling repetition and exploration in the recommended basket. 
Besides, we also think that future NBR methods should be able to combine repetition and exploration based on users' preferences.

\subsection{Limitations}
One of the limitations of this study is that we did not consider the training and inference execution time in the paper, which is important for the real-world value of methods used for \ac{NBR}~\citep{ariannezhad-2023-personalized}. We use the original implementations of \acs{NBR} methods to check their reproducibility and avoid potential mistakes that may come with re-implementations, however, the original implementations are based on different frameworks, which leads to an inability to make a fair execution time comparison.
A second limitation is that we only follow the widely used binary definition of repeat items and explore items but do not consider a more fine-grained formalization based on the frequency of historical items, which would allow for a more flexible definition of repetition and analysis.
A further limitation is that we only considered the short-term utility of \ac{NBR} methods: will users be satisfied with their next basket? Limited by our experimental setup, where we replay users' past behavior, we have ignored any potential long-term effects of having a strong focus on short-term utility by emphasizing repeat items as opposed to, for instance, long-term engagement which, likely, benefits from a certain degree of exploration so as to enable surprise and discovery.

\subsection{Future work}
Obvious avenues for future work include addressing the limitations that we have summarized above. 
Another important line of future work concerns the use of domain-specific knowledge, either concerning complementarity or substitutability of items or concerning hierarchical relations between items, both of which would allow one to consider more semantically informed notions of repeat consumption behavior~\citep{ariannezhad-2022-recanet} for \acl{NBR} purposes. 
In addition, our focus in this paper has been on users -- in the sense that we compared methods that produce a basket for a given user --, it would be interesting to consider repetition and exploration aspects of the reverse scenario~\citep{li-2023-who} -- given an item, who are the users to whose baskets this item can best be added?
Finally, even though we focused on \acl{NBR}, it would be interesting to contrast our outcomes with an analysis of repeat and explore behavior in traditional sequential recommendation scenarios. 

\section*{Supplementary materials}
To facilitate reproducibility of our work, we share the following materials at \url{https://github.com/liming-7/A-Next-Basket-Recommendation-Reality-Check}):
\begin{enumerate*}[label=(\roman*)]
\item source code and datasets;
\item descriptions of different dataset formats; and
\item pipelines for running our experiments and obtaining our results.
\end{enumerate*}

\begin{acks}
We would like to thank our reviewers and associate editor for their constructive feedback.

This research was partially supported by Ahold Delhaize,
the China Scholarship Council (grant \#20190607154),
and the Hybrid Intelligence Center, a 10-year program funded by the Dutch Ministry of Education, Culture and Science through the Netherlands Organisation for Scientific Research, \url{https://hybrid-intelligence-centre.nl}.

All content represents the opinion of the authors, which is not necessarily shared or endorsed by their respective employers and/or sponsors.
\end{acks}

\bibliographystyle{ACM-Reference-Format}
\bibliography{references}


\appendix

\section{Appendix}
\subsection{Additional plots}

\begin{figure}[h]
    \begin{subfigure}{\textwidth}
    \centering
    \includegraphics[width=\textwidth]{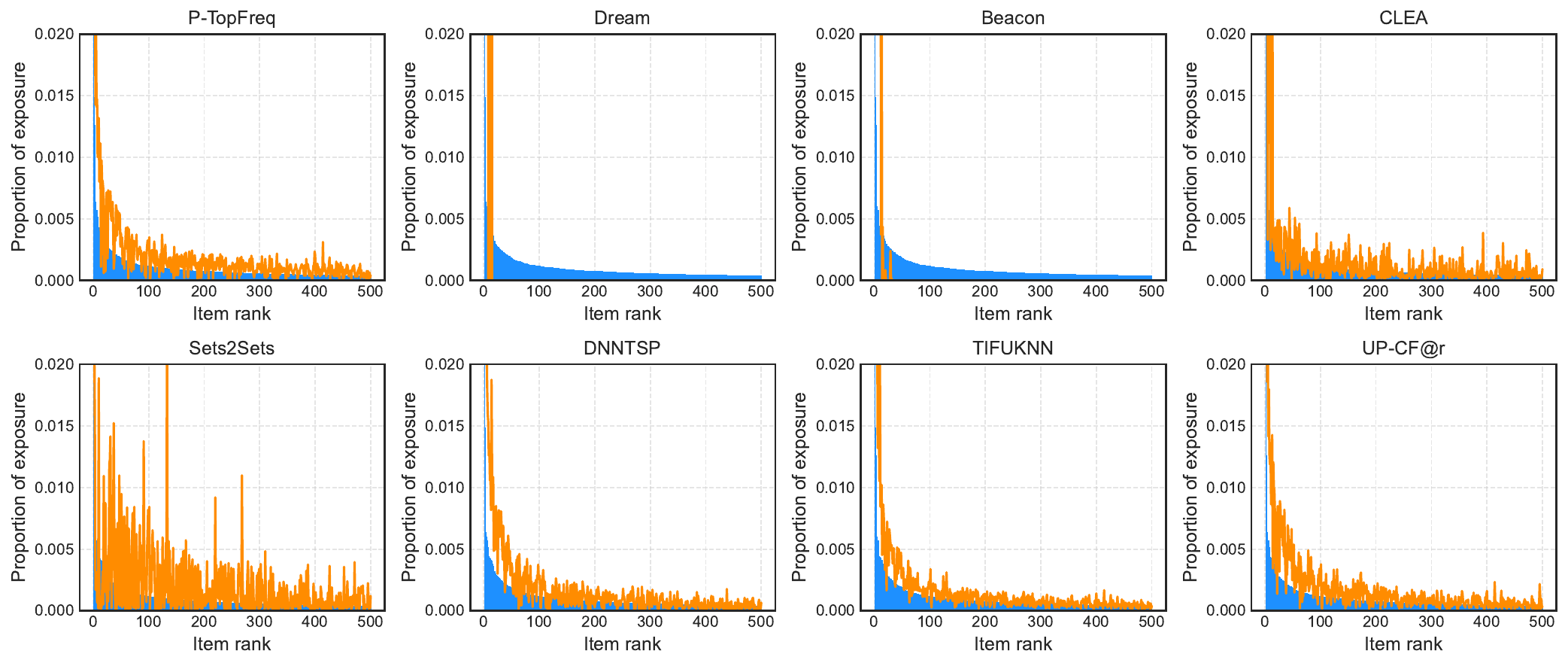}
    \caption{Items ranked according to their frequencies in the training labels.}
    \label{fig:dunn_exposure_labels}
    \end{subfigure}
    
    \vspace*{2mm}
    \begin{subfigure}{\textwidth}
    \centering
    \includegraphics[width=\textwidth]{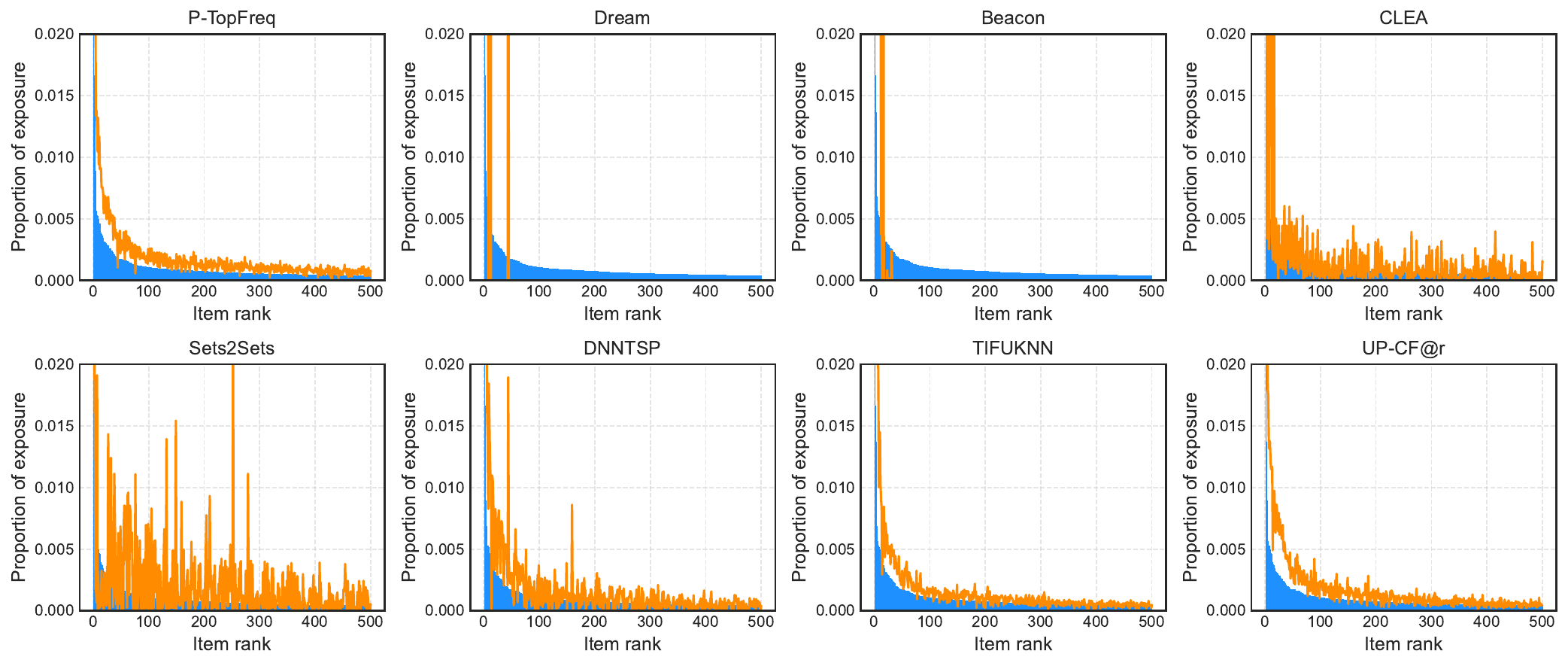}
    \caption{Items ranked according to their frequencies in the testing inputs (i.e., historical baskets of testing users).}
    \label{fig:dunn_exposure_historical}
    \end{subfigure}
    \caption{Treatment effect for items on the Dunnhumby dataset, for eight \ac{NBR} methods for items with different frequencies in training labels and testing inputs. The blue bar shows the frequency distribution, and the orange line denotes the exposure distribution.}
    \label{fig:dunn_item_level_exposure_analysis}
  \end{figure}

\begin{figure}[h]
    \begin{subfigure}{\textwidth}
    \centering
    \includegraphics[width=\textwidth]{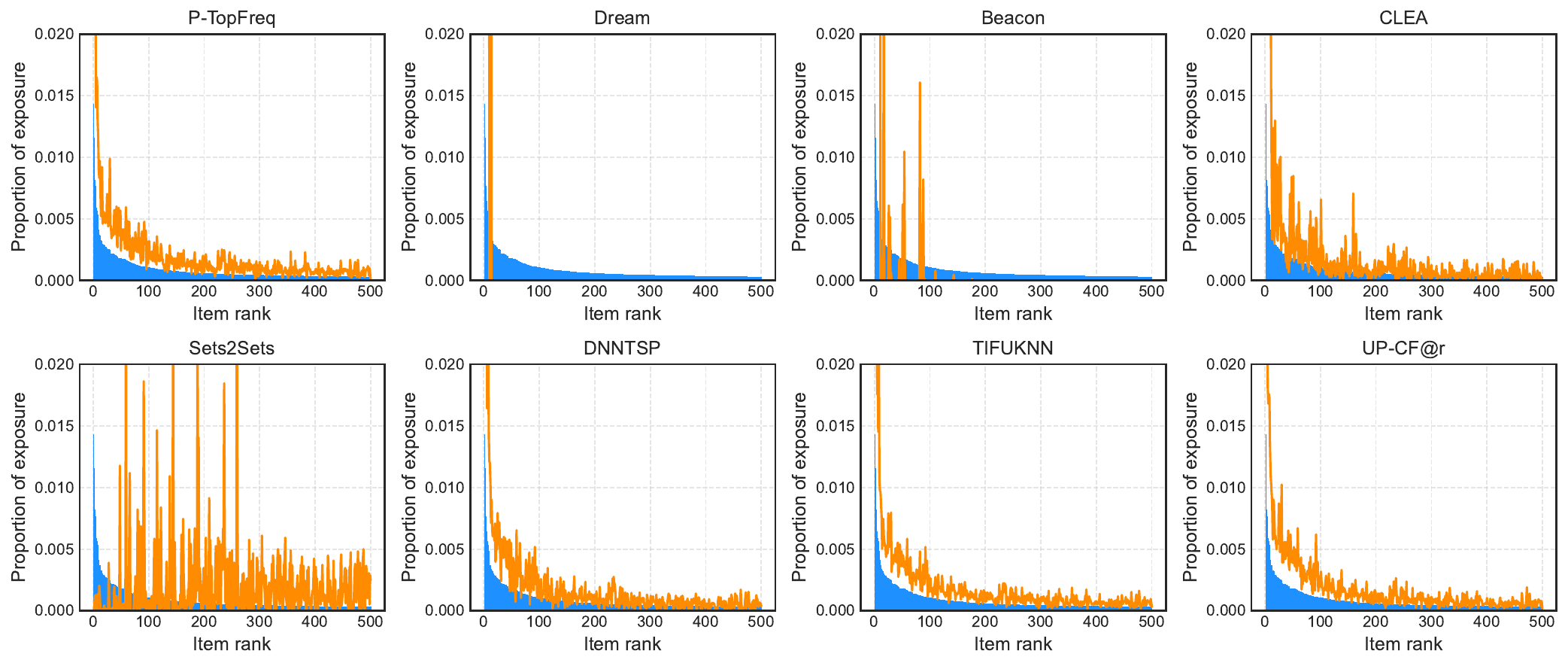}
    \caption{Items ranked according to their frequencies in the training labels.}
    \label{fig:inst_exposure_labels}
    \end{subfigure}
    
    \vspace*{2mm}
    \begin{subfigure}{\textwidth}
    \centering
    \includegraphics[width=\textwidth]{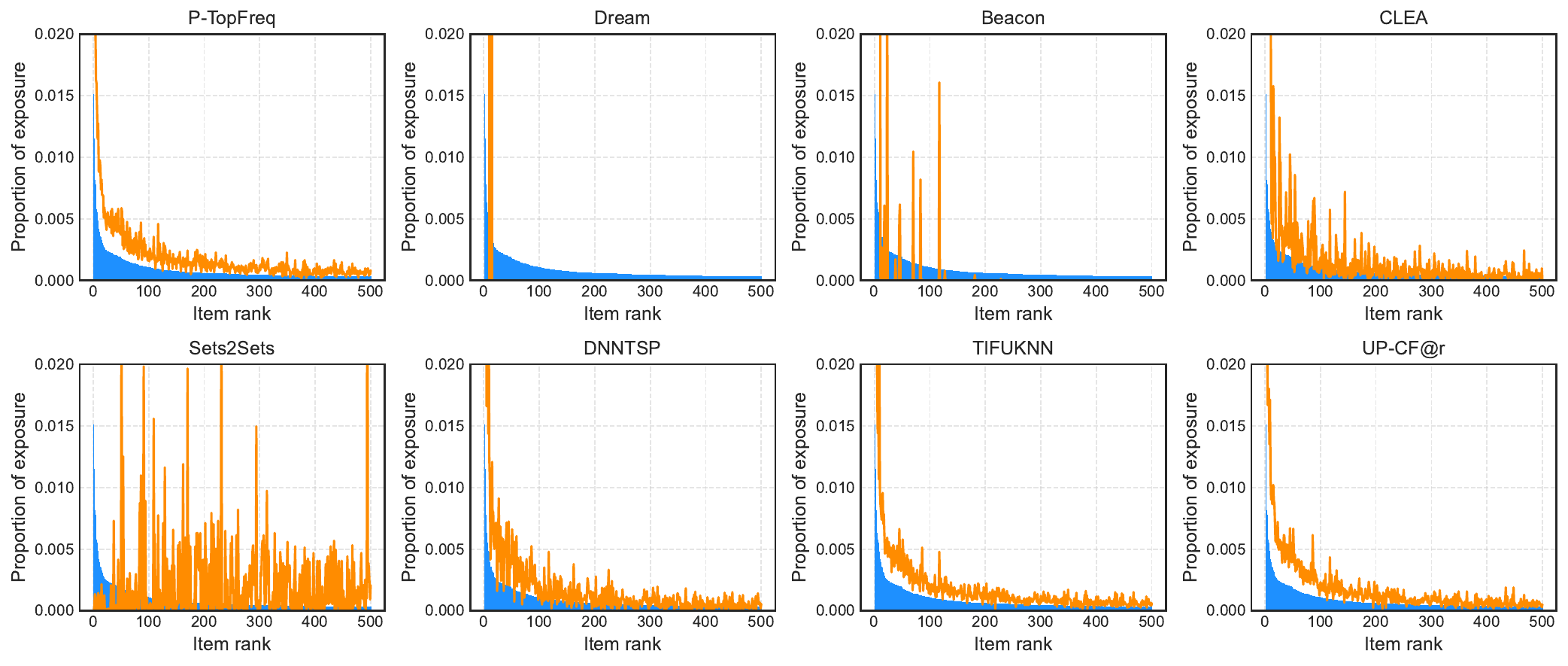}
    \caption{Items ranked according to their frequencies in the testing inputs (i.e., historical baskets of testing users).}
    \label{fig:inst_exposure_historical}
    \end{subfigure}
    \caption{Treatment effect for items on the Instacart dataset, for eight \ac{NBR} methods for items with different frequencies in training labels and testing inputs. The blue bar shows the frequency distribution, and the orange line denotes the exposure distribution.}
    \label{fig:inst_item_level_exposure_analysis}
  \end{figure}

We include additional plots to supplement Figure~\ref{fig:repr_distribution} (Figure~\ref{fig:repr_distribution_app} below) and~\ref{fig:item_level_exposure_analysis} (Figures~\ref{fig:dunn_item_level_exposure_analysis} and~\ref{fig:inst_item_level_exposure_analysis}), respectively.

\begin{figure}[h]
\begin{subfigure}{\textwidth}
    \centering
    \includegraphics[width=\textwidth]{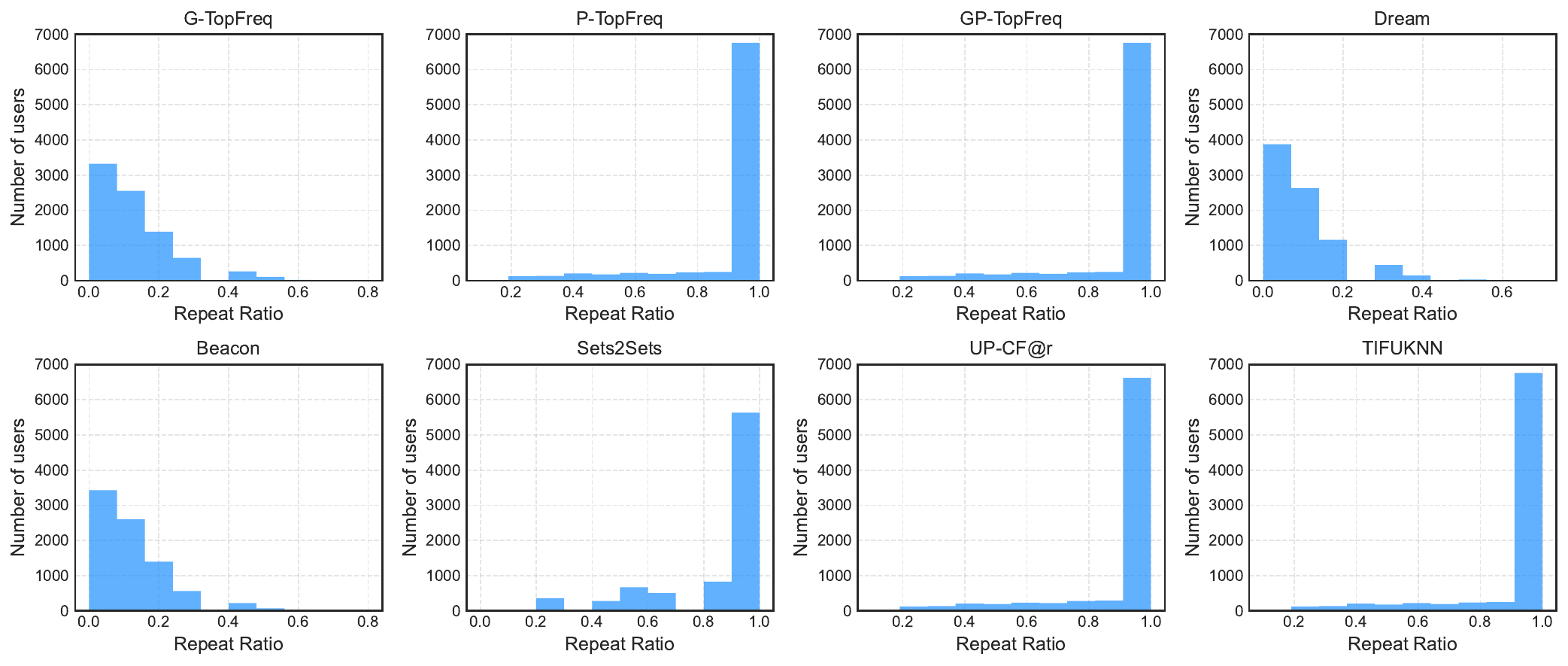}
    \caption{TaFeng}
    \label{fig:tafeng_repr_distribution_app}
\end{subfigure}
\begin{subfigure}{\textwidth}
    \centering
    \includegraphics[width=\textwidth]{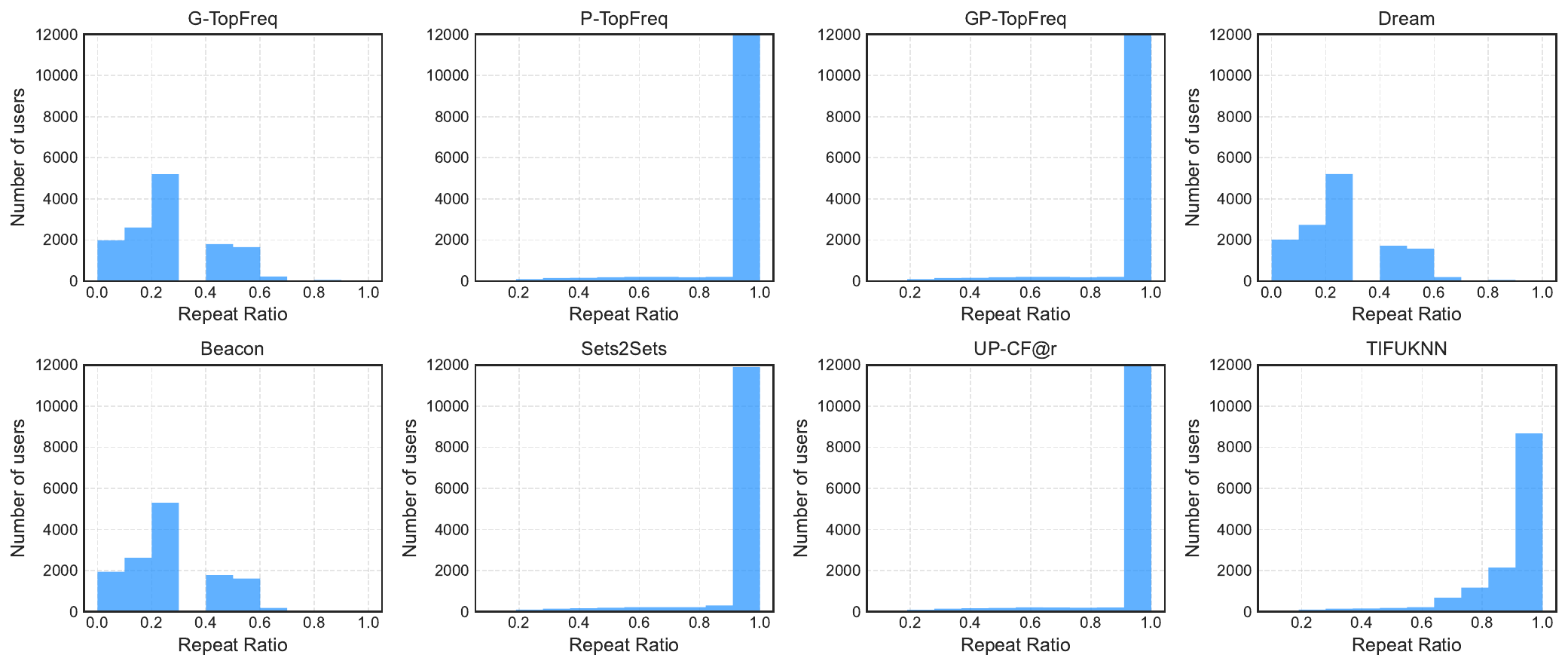}
    \caption{Dunnhumby}
    \label{fig:dunnhumby_repr_distribution_app}
\end{subfigure}
\begin{subfigure}{\textwidth}
    \centering
    \includegraphics[width=\textwidth]{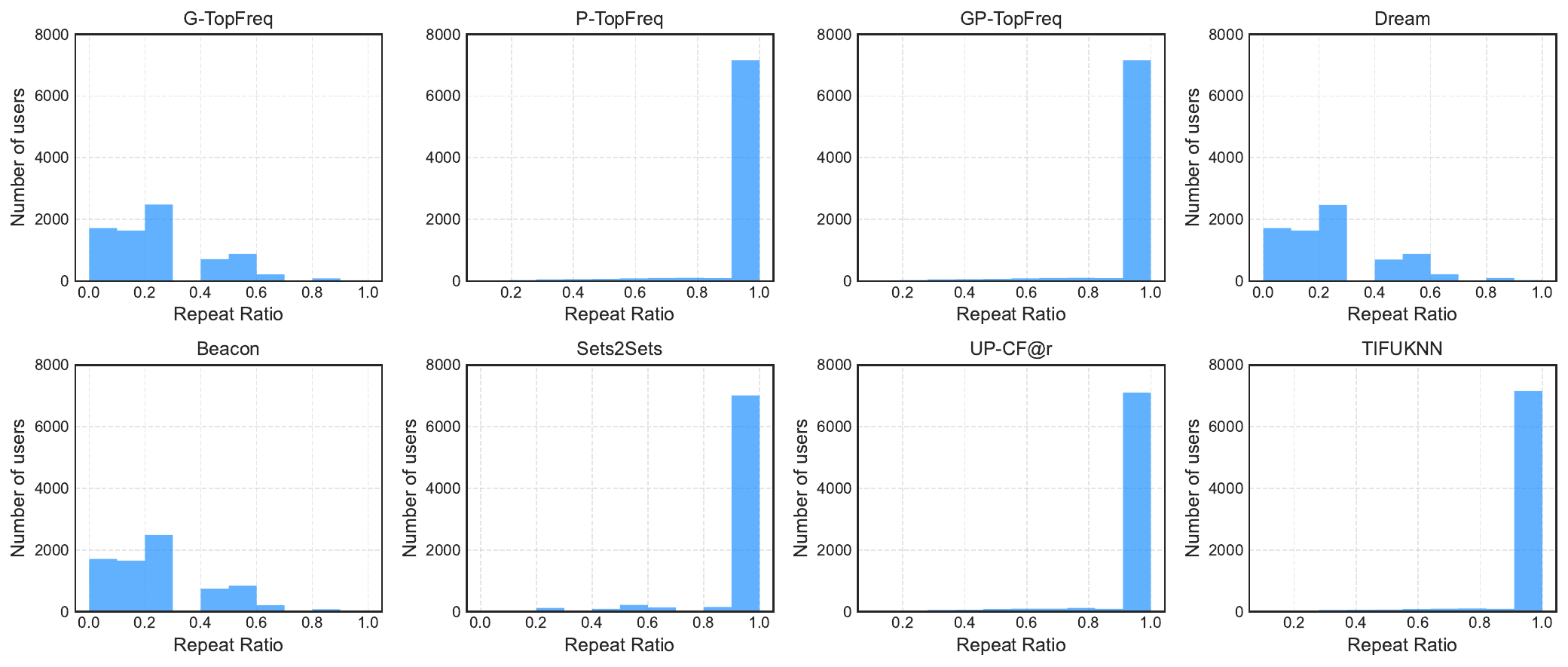}
    \caption{Instacart}
    \label{fig:instacart_repr_distribution_app}
\end{subfigure}
\caption{Distribution of the repetition ratio $\RepR$ of recommended baskets on TaFeng dataset for eight NBR methods (G-TopFreq, P-TopFreq, GP-TopFreq, Dream, Beacon, Sets2Sets, Up-CF@r, TIFUKNN).}
\label{fig:repr_distribution_app}
\end{figure}

\subsection{Reproducibility}
To facilitate reproducibility of the results in this paper, our online repository, which can be found at \url{https://github.com/liming-7/A-Next-Basket-Recommendation-Reality-Check}, contains the following resources:
\begin{enumerate*}[label=(\roman*)]
\item source code and datasets;
\item descriptions of different dataset format; and
\item pipelines about how to run and get results.
\end{enumerate*}

\end{document}